\documentclass[12pt]{article}
\usepackage{graphicx}
\usepackage[english]{babel}
\usepackage{amsmath}
\usepackage[labelfont=bf]{caption}

\usepackage{scicite}

\usepackage{times}

\newenvironment{sciabstract}{%
\begin{quote} \bf}
{\end{quote}}

\newcounter{lastnote}

\renewcommand{\refname}{References and Notes}

\title{Magnetic moment and plasma environment of HD 209458b as determined from Ly$\alpha$ observations$^{**}$}

\author
{Kristina G. Kislyakova,$^{1\ast}$ Mats Holmstr{\"o}m,$^{2}$ Helmut Lammer,$^{1}$  \\
Petra Odert,$^{3}$ Maxim L. Khodachenko$^{1,4}$ \\
\\
\normalsize{$^{1}$Space Research Institute, Austrian Academy of Sciences,}\\
\normalsize{Schmiedlstrasse 6, A-8042 Graz, Austria}\\
\normalsize{$^{2}$Swedish Institute of Space Physics,}\\
\normalsize{PO Box 812, SE-98128 Kiruna, Sweden}\\
\normalsize{$^{3}$Institute of Physics, University of Graz,}\\
\normalsize{Universit{\"a}tsplatz 5, A-8010 Graz, Austria}\\
\normalsize{$^{4}$Skobeltsyn Institute of Nuclear Physics, Moscow State University,}\\
\normalsize{Leninskie Gory, 119992 Moscow, Russia}\\
\\
\normalsize{$^\ast$To whom correspondence should be addressed. E-mail:  kristina.kislyakova@oeaw.ac.at}
}

\date{}

\begin{document} 


\baselineskip24pt


\maketitle 

\textbf{$^{**}$ This is the author's version of the work. It is posted here by permission of the AAAS for non-commercial research use only. The definitive version was published in \textit{Science} (Vol. \textbf{346}, p. 981, 21 November 2014), DOI: 10.1126/science.1257829 }


\begin{sciabstract}
Transit observations of HD 209458b in the stellar Lyman-$\alpha$ (Ly$\alpha$) line revealed strong absorption in both blue and red wings of the line interpreted as hydrogen atoms escaping from the planet's exosphere at high velocities. The following sources for the absorption were suggested: acceleration by the stellar radiation pressure, natural spectral line broadening, charge exchange with stellar wind. 
We reproduce the observation by means of modelling that includes all aforementioned processes. Our results support a stellar wind with a velocity of $\approx400$~km$\times$s$^{-1}$ at the time of the observation and a planetary magnetic moment of $\approx 1.6 \times 10^{26}$~A$\times$m$^2$.
\end{sciabstract}

% In setting up this template for *Science* papers, we've used both
% the \section* command and the \paragraph* command for topical
% divisions.  Which you use will of course depend on the type of paper
% you're writing.  Review Articles tend to have displayed headings, for
% which \section* is more appropriate; Research Articles, when they have
% formal topical divisions at all, tend to signal them with bold text
% that runs into the paragraph, for which \paragraph* is the right
% choice.  Either way, use the asterisk (*) modifier, as shown, to

%\section*{Introduction}

%HD~209458b with a mass of $M_{\rm pl} = 0.71 M_{\rm Jup}$ and a radius of $R_{\rm pl} = 1.38 R_{\rm Jup}$ orbits its star at $\approx 0.047$~AU. HD~209458 is a solar-type G star with the mass of $M_{\rm st} = 1.148 M_{\rm Sun}$ and the age of $\sim$4~Gyr\cite{exoplanet.eu}. 
Hubble Space Telescope observations revealed a strong Ly$\alpha$ absorption of 15$\pm$4 \% (1$\sigma$) during the transit of the exoplanet HD 209458b	 in front of its host star. This Ly$\alpha$ absorption significantly exceeded the 1.5\% absorption in the visible\cite{VM03}. Shortly afterward, the data were reanalyzed, and a lower absorption depth of 8.9$\pm$2.1\% was reported\cite{BJ07}, still present in both blue and red wings of the Ly$\alpha$ line. The frequency $f$ and velocity spectrum of neutral H atoms along the line-of-sight (LOS) $v_x$ are connected through the relation 
\begin{equation}
	f = f_0 + v_x/\lambda_0,
	\label{e_vel}
\end{equation}
where $\lambda_0 = 1215.65 \times 10^{-10}$~m. Absorption in the wings of the Ly$\alpha$ line is a signature of neutral hydrogen atoms moving with high velocities towards (positive velocities along the LOS) and away from the star. Several interpretations for the excess absorption have been suggested: escaping neutral hydrogen atoms accelerated by the stellar radiation pressure\cite{VM03}, natural spectral line broadening\cite{BJ10}, and energetic neutral atoms (ENAs)\cite{H08,E10,Tremblin13}.

%---------------------------------------------------------------------
\begin{table}
\begin{minipage}{\textwidth}
  \caption[]{Parameters of HD209458 and HD209458b\cite{Southworth10} and some simulation parameters.}
  $$ 
  \begin{array}{p{0.5\linewidth}lll}
    \hline
    \noalign{\smallskip}
    Name &  \mathrm{Symbol} & \mathrm{Value} \\
    \noalign{\smallskip}
    \hline
    \noalign{\smallskip}
Star mass		 	& M_{\rm st} 	& 2.28\times 10^{30}~\mathrm{kg} & \approx 1.148~M_{\rm Sun}\\
Star age		 	&  	 	& \approx 4\pm2~{\rm Gyr}  &\\
Planet radius 		 	& R_{\rm pl}	& 9.54\times 10^7~\mathrm{m} & \approx 0.71~M_{\rm Jup} \\
Planet mass 		 	& M_{\rm pl}	& 1.21\times 10^{27}~\mathrm{kg} &\approx 1.38~R_{\rm Jup}\\
Orbital distance	 	& 		& 7.1\times 10^9~\mathrm{m} & \approx 0.047~\mathrm{AU}\\
Inner boundary radius    	& R_{\rm ib}	& 2.7\times 10^8~\mathrm{m} & \approx 2.8~R_{\rm pl} \\
Inner boundary temperature	& T_{\rm ib}	& 6\times 10^3~\mathrm{K} & \\
Inner boundary density 		& n_{\rm ib}		& 2\times 10^{13}~\mathrm{m}^{-3}& \\
Obstacle standoff distance$^a$	& R_{\rm s} 	 	      & 2.76 \times 10^8 \mathrm{m} & \approx 2.9~R_{\rm pl} \\
Obstacle width\footnote{In the assumption that the Alfvenic Mach number $M_A > 1$}		& R_{\rm t} 	  	      & 2.86 \times 10^8 \mathrm{m} & \approx 3.0~R_{\rm pl} \\
Photoionization rate 		& \tau_{\mathrm{pi}}  & 6.0\times 10^{-5}~\mathrm{s}^{-1} & \\
Electron impact ionization rate	& \tau_{\mathrm{ei}}  & 1.25\times 10^{-4}~\mathrm{s}^{-1} & \\
Stellar wind density		& n_{\rm sw}     & 5\times 10^9~\mathrm{m}^{-3} & \\
Stellar wind velocity		& u_{\rm sw}     & 400 \times 10^3~\mathrm{m/s} & \\
Stellar wind temperature		& T_{\rm sw}     & 1.1 \times 10^6~\mathrm{K} & \\
    \noalign{\smallskip}
    \hline
  \end{array}
  $$ 
\label{t_par}
\end{minipage}
\end{table}
%---------------------------------------------------------------------

In the first scenario, neutral atoms are accelerated by the radiation pressure of stellar Ly$\alpha$ photons until the acceleration force is balanced by the gravity of the star. 

Spectral line broadening arises due to the natural broadening mechanism and the spread of the particle velocities along the LOS. One must also account for the populations of non-thermal hydrogen accelerated by the radiation pressure and formed by charge exchange. Their broadening is non-symmetric and is not described by the Voigt profile, which is the convolution of the Lorentz and Doppler profiles, as is the case for purely thermal populations.

ENAs are generally produced by charge exchange between ions and neutral atoms. In the exosphere, charge exchange occurs between high-velocity stellar wind protons and neutral atmospheric particles moving at thermal velocities. In this interaction, the electron is transferred to the proton. Because the initial velocities of the particles are almost conserved, the reaction produces a "slow" ion and a "fast" neutral. This mechanism generates ENA coronae around the Earth and other Solar system planets\cite{Fahr07} and should operate in exoplanet systems as well\cite{H08}.

None of the early studies considered all these effects together. A later work\cite{BL13} included all processes, but was still unable to explain the absorption in the red part of the line, which was recently observed again also for another exoplanet GJ~436b\cite{Kulow2014}. The aforementioned work\cite{BL13} includes simplified natural broadening model. They use a Lorentzian broadening profile, that is derived from the natural broadening of a Maxwellian gas. However, here we have a non-Maxwellian population of hydrogen atoms. They also use a too coarse velocity grid when computing the naturally broadened profile. The computation has to be done on a grid that has much higher resolution to be accurate.

Here we show that the red part of the absorption can be explained by spectral broadening alone, while the blue part also reflects contributions from radiation pressure and ENAs.

%\section*{Method Description}

With numerical modelling, we estimate how a neutral hydrogen cloud located in front of the star would affect the transit Ly$\alpha$ absorption of HD~209458b. Parameters of the system are summarized in Table~\ref{t_par}. The orbit of HD~209458b is almost circular. The simulations were performed by using a Direct Simulation Monte-Carlo code, which includes neutral hydrogen atoms and hydrogen ions. The comprehensive description of the model can be found in \cite{SM}. The main processes and forces included for an exospheric atom are:
\begin{enumerate}
 \item Collision with a UV photon which defines the velocity-dependent radiation pressure, self-shielding\cite{BL13} is included.
 \item Charge exchange with a stellar wind proton.
 \item Elastic collision with another hydrogen atom.
 \item Ionization by stellar photons or wind electrons. For a solar-type star, at an orbit of 0.047~AU, the electron impact ionization rate exceeds the photoionization rate by a factor of two (Table \ref{t_par}) due to a denser stellar wind and thus cannot be neglected\cite{Holzer77}.
 \item Gravity of the star and planet, centrifugal, Coriolis and tidal forces.
\end{enumerate}

In this study we use a velocity-dependent absorption coefficient of photons\cite{K14}. The photon absorption rate has the same shape as the stellar Ly$\alpha$ line, so that the strongest acceleration of neutral hydrogen atoms occurs in the velocity domain between approximately $-100\le v_x \le 100$~km/s. Self-shielding\cite{BL13} is also included: in the regions optically thick in Ly$\alpha$ the neutrals cannot collide with UV photons. 

After the hydrogen corona is modelled, the Ly$\alpha$ in-transit attenuation is calculated. Natural broadening is accounted for at this stage. Doppler broadening is included automatically by accurately taking into account the velocities of the atoms.

Since HD~209458 is a solar-type star, we assume the solar parameters at 0.047~AU for stellar wind density, temperature and velocity and use scaled photo- and electron impact ionization rates from the Sun\cite{Holzer77}.

%\section*{Results}

Our results show that an extended hydrogen corona exists around HD~209458b. We expect this to form when strong radiation pressure accelerates neutral hydrogen atoms and moves them out of the planetary magnetosphere, where they may undergo ionization either by charge exchange or by stellar wind electrons. Photoionization can occur in the whole non-shielded atmosphere outside the planet's shadow. The non-symmetric form of the corona is defined by radiation pressure, charge exchange and the Coriolis force. Our results show that 1D atmospheric modelling\cite{BJ10} that assumes a symmetric upper atmosphere, should be applied cautiously above the exobase to highly irradiated exoplanets.

%------------------------------------------------------------------
%\begin{figure}[htb!]
\begin{figure}[ht]
  \centering
  \includegraphics[width=1.0\textwidth]{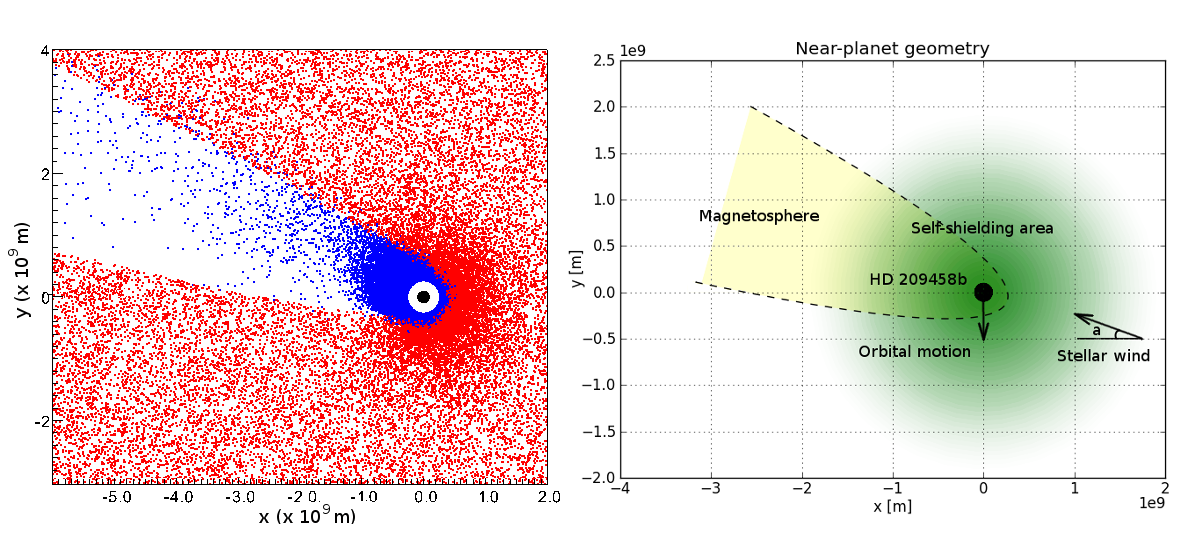}
  \caption{Left panel: slice of modelled 3D atomic hydrogen corona around HD~209458b. Blue and red dots correspond to neutral hydrogen atoms and hydrogen ions including stellar wind protons, respectively. The black dot represents the planet. The white area around the planet corresponds to the thermosphere below the inner simulation boundary at the height $R_{\rm ib}$. The star is on the right. The planetary magnetosphere is represented by an obstacle with a width $R_{\rm t}$ and a distance from the planet's center to substellar point $R_{\rm s}$. We assume the atmosphere of HD~209458b to be dominated by atomic hydrogen at $R_{\rm ib}$\cite{Koskinen13,Shematovich10}. Hydrogen atoms are launched from the inner boundary assuming a number density $n_{\rm ib}$ and a temperature of $T_{\rm ib}$, consistent with atmospheric models\cite{Koskinen13,Koskinen10}. Stellar wind protons are launched on the right side of the domain and cannot penetrate inside the magnetospheric obstacle. Right panel: llustration of near-planet geometry.}
\end{figure}
%-----------------------------------------------------------------

Fig.~1 illustrates the modelled hydrogen cloud around HD~209458b. The coordinate system is centered at the center of the planet, the $x$-axis points towards the center of mass of the system, the $y$-axis points in the opposite direction to the planet's velocity and the $z$-axis completes the right-hand coordinate system. Radiation pressure together with the tidal force lead to a formation of a cometary-like tail\cite{BL13,DAngelo14}. One can clearly see the interaction area at the magnetospheric boundary revealed by enhanced concentration of H$^+$ ions (red dots), where the neutrals underwent charge exchange. Fig.~2 presents the observed\cite{BJ10} and modelled Ly$\alpha$ spectra with and without natural broadening normalized by the out-of-transit observation. Broadening plays the key role in absorption in the red wing of the line (positive $v_{\rm x}$). This conclusion is in agreement with an early study\cite{BJ10}, which, however, disregarded ENAs and radiation pressure contributing to absorption. Additionally, it was based on a 1D symmetric model of the atmosphere, while the modelled corona around HD~209458b shows significant asymmetry.

%------------------------------------------------------------------
%\begin{figure}[htb!]
\begin{figure}[ht]
  \centering
  \includegraphics[width=0.7\textwidth]{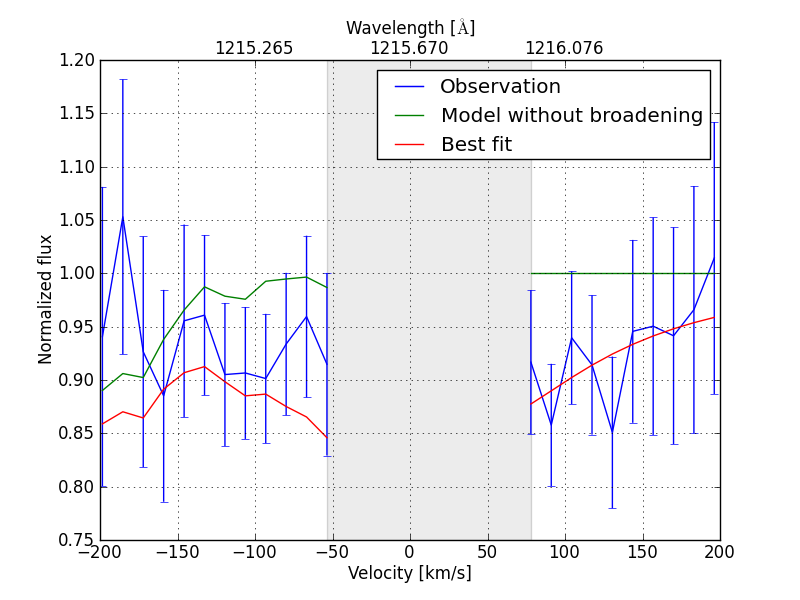}
  \caption{Comparison of modelled and observed (according to Ben-Jaffel and Hosseini\cite{BJ10}) Ly$\alpha$ spectra at mid-transit. Blue, the observed in-transit profile, normalized by the out-of-transit profile. Red, the modelled profile, computed including charge exchange, radiation pressure with self-shielding, photo- and electron impact ionization, and natural broadening. The goodness of the best fit equals $\chi^2 = 0.081$. Green, the same as red, but excluding broadening, $\chi^2 = 0.139$. The abscissa is the hydrogen velocity along the $x$ axis (away from Earth, towards the star). The region of contamination by geocoronal emission at low velocities is excluded and marked by the shaded area. On can clearly see that broadening affects both wings of the Ly$\alpha$ line, while other effects contribute to the absorption mainly on the blue wing produced by atoms moving away from the star. The details of computing the Ly$\alpha$ attenuation are given in \cite{SM}.}
\end{figure}
%-----------------------------------------------------------------

 Finally, Fig.~3 depicts the velocity spectrum of neutral hydrogen atoms along the star-planet line with two populations of atoms: the left peak shows the ENAs, the right higher peak the thermal atmospheric atoms. 
 
%------------------------------------------------------------------
%\begin{figure}[htb!]
\begin{figure}[ht]
  \centering
  \includegraphics[width=0.7\textwidth]{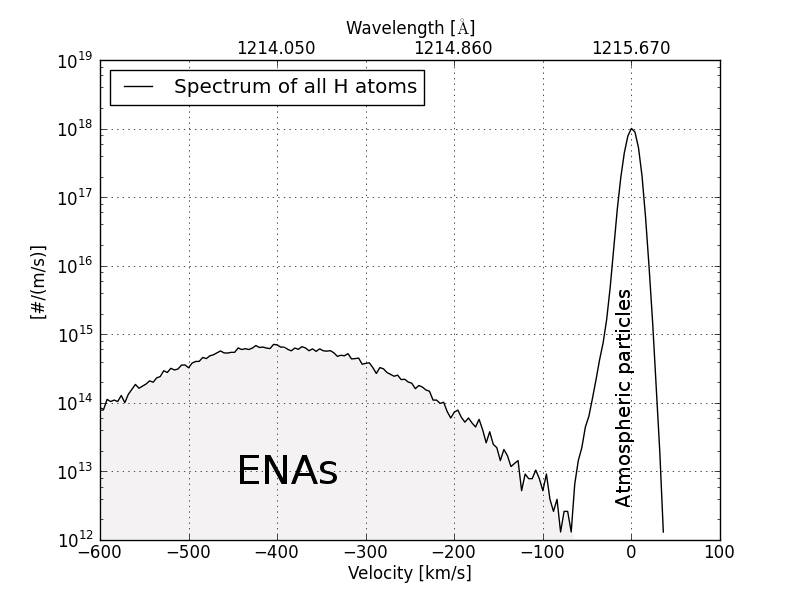}
  \caption{The modelled best-fit $x$-axis (planet-star) velocity spectrum of neutral hydrogen atoms in front of the star at mid-transit. One can see two hydrogen populations: the thermal population of the atmosphere (right peak) and the ENAs produced by charge exchange (left peak). The width of this part of the distribution is proportional to the stellar wind temperature. Note that the thermal peak is slightly deformed and shifted towards negative velocities by the radiation pressure.}
\end{figure}
%-----------------------------------------------------------------
 
 In the velocity spectrum corresponding to our best fit, the ENA peak is located near $-400$~km/s due to a stellar wind (Fig.~3). We do not take into account the effects connected with the formation of a bow shock near the magnetospheric obstacle in the supersonic regime. In both cases when the bow shock is present or absent, there exist an ENA formation region. It is not easy to distinguish if the estimated plasma parameters are from the undisturbed stellar wind or the slower and hotter shocked stellar wind. However, the best estimate we have of a 400~km/s indicates that the ENAs are produced in an undisturbed (possibly deviated, but not shocked) stellar wind. The observation cannot be fitted with lower stellar wind velocities for any plausible planetary magnetic field \cite{SM}. Because the ENAs maintain the velocity of the protons, fast stellar wind shifts their location out of the considered velocity range ($-200$~km/s$\le v_{\rm x} \le 200$~km/s, because the observed signal becomes too noisy beyond, see Fig.~2) to avoid overabsorption. However, ENAs play an important role in reducing the amount of neutrals in the velocity domain of interest, shaping the cloud and contributing to the absorption in the very blue part of the spectrum. Our search of realistic parameter space led us to conclude that the observations can be fitted only by using a very narrow and close-in magnetic obstacle with a sub-stellar point located at $\approx 2.9 R_{\rm pl}$ from the planet's center and an obstacle width of $\approx 3 R_{\rm pl}$. In the case of a close-in obstacle, the exosphere interacts directly with the stellar wind. This diminishes the number of hydrogen neutrals, first, by charge exchange and later by stellar wind electron impact ionization. A stronger intrinsic field shifts the magnetic boundary away from the planet and effectively protects the atmosphere from these processes. This dramatically increases the number of neutrals which undergo acceleration by radiation pressure up to $-200$~km/s and leads to overabsorption in comparison to the in-transit observation. One should also state we could only fit the observations by including the self-shielding effect\cite{BL13,SM}.
 
Assuming a dipole approximation, one can estimate the magnetic moment of HD~209458b using the formula\cite{Khodachenko12}
 
\begin{equation}
	\mathcal{M} = \left( \frac{8 \pi^2 R_{\rm s}^6 \rho_{\rm sw} v_{\rm rel}^2}{\mu_0 f_0^2} \right)^{1/2}
	\label{e_moment}
\end{equation}
Here, $R_{\rm s}$ is the magnetospheric stand-off distance, $\mu_0$ is the diamagnetic permeability of free space, $f_0 \approx 1.22$ is a form factor of the magnetosphere, $\rho_{\rm sw}$ is the mass density of the stellar wind, and $v_{\rm rel}$ is the relative velocity of the stellar wind plasma, including the planetary orbital velocity.

Adopting our best-fit simulation parameters, Eq.~\ref{e_moment} yields $\mathcal{M} \approx 1.6 \times 10^{26}$~A$\times$m$^2 \approx 0.1~\mathcal{M}_{\rm Jup}$ for HD~209458b, where $\mathcal{M}_{\rm Jup}=1.56\times 10^{27}$~A$\times$m$^2$ \cite{Griessmeier04}. This value corresponds to the upper boundary predicted by the models of tidally locked close-in exoplanets\cite{Griessmeier04, Sanchez-Lavega04}, atmospheric models\cite{Koskinen10} and an empirical relationship\cite{Durand-Manterola09} and is in agreement with the non-detection of the radio emission from HD~209458b at 150 MHz\cite{Lecavelier11}. Proximity of this value to a dipole field approximation indicates a negligible magnetodisk contribution, which is in agreement with the conclusion that HD~209458b probably can not develop a significant magnetodisk\cite{Khodachenko12}. Our results do not support a larger magnetic moment that exceeds $\mathcal{M}_{\rm Jup}$, as predicted by an energy flux scaling model\cite{Christensen09,Reiners10}.

In summary, our model predicts a fast stellar wind at the time of observation ($\approx 400$~km/s, $n_{\rm sw} \approx 5 \times 10^9$~m$^{-3}$), stresses the importance of electron impact ionization in addition to photoionization, and infers a magnetic moment of HD~209458b of $\approx$10\%~$\mathcal{M}_{\rm Jup}$.

% Your references go at the end of the main text, and before the
% figures.  For this document we've used BibTeX, the .bib file
% scibib.bib, and the .bst file Science.bst.  The package scicite.sty
% was included to format the reference numbers according to *Science*
% style.

\renewcommand{\refname}{References and Notes}
\nocite{*} %to include references to ALL citations in the references_HD209458b.bib
%\bibliography{scibib}
\bibliography{references_HD209458b}
\bibliographystyle{Science}

% Following is a new environment, {scilastnote}, that's defined in the
% preamble and that allows authors to add a reference at the end of the
% list that's not signaled in the text; such references are used in
% *Science* for acknowledgments of funding, help, etc.

%\begin{scilastnote}
\textbf{Acknowledgements.} This study was carried out with the support by the FWF NFN project S116601-N16 ''Pathways to Habitability: From Disk  to Active Stars, Planets and Life'' and the related FWF NFN subprojects S116 606-N16, and S116 607-N16. P. Odert thanks Austrian Science Fund (FWF): P22950-N16. The research was conducted using computer resources provided by the Swedish National Infrastructure for Computing (SNIC) at the High Performance Computing Center North (HPC2N). The software was in part developed by the DOE NNSA-ASC OASCR Flash Center at the University of Chicago. The authors thank L. Ben-Jaffel for providing the reprocessed HST observations, B. Wood for providing the processed Ly$\alpha$ profile, the ISSI team "Characterizing stellar and exoplanetary environments" for useful discussions, and C.P.~Johnstone for his valuable comments on the stellar wind issue. The data are available in Supporting Online Material.
%\end{scilastnote}

\textbf{Supplementary Content.}\newline
Supplementary Text\newline
Figs. S1 to S12\newline
Tables S1 to S2\newline
References (25 -- 42)

\newpage
\section*{Supplementary Materials for Magnetic Moment And Plasma Environment Of HD 209458b As Determined From Ly$\alpha$~Observations}
%\end{center}

\textbf{Published in \textit{Science} (Vol. \textbf{346}, p. 981, 21 November 2014), DOI: 10.1126/science.1257829}

\renewcommand{\thefigure}{S\arabic{figure}}
\renewcommand{\thetable}{S\arabic{table}}
\setcounter{figure}{0}
\setcounter{equation}{0}

{Kristina~G.~Kislyakova,$^{1\ast}$ Mats~Holmstr{\"o}m,$^{2}$ Helmut~Lammer,$^{1}$  Petra~Odert,$^{3}$ Maxim~L.~Khodachenko$^{1,4}$ \\
\\
\normalsize{$^{1}$Space Research Institute, Austrian Academy of Sciences,}\\
\normalsize{Schmiedlstrasse 6, A-8042 Graz, Austria}\\
\normalsize{$^{2}$Swedish Institute of Space Physics,}\\
\normalsize{PO Box 812, SE-98128 Kiruna, Sweden}\\
\normalsize{$^{3}$Institute of Physics, University of Graz,}\\
\normalsize{Universit{\"a}tsplatz 5, A-8010 Graz, Austria}\\
\normalsize{$^{4}$Skobeltsyn Institute of Nuclear Physics, Moscow State University,}\\
\normalsize{Leninskie Gory, 119992 Moscow, Russia}\\
\\
\normalsize{Correspondence to:  kristina.kislyakova@oeaw.ac.at.}
}
~\newline
~\newline
\textbf{This PDF file includes:}

Supplementary Text

Figs. S1 to S12

Tables S1 to S2

In this section we describe the numerical algorithms used for the modelling of the hydrogen exosphere around HD~209458b and discuss in more detail some relevant physical effects and the influence of changing the parameters.
~\newline
\subsection*{The Simulation Algorithm}
We model the plasma interaction between the stellar wind and the upper atmosphere of HD~209458b by applying a Direct Simulation Monte Carlo (DSMC) upper atmosphere-exosphere 3D particle model. The software is based on the FLASH code written in Fortran 90, which was developed at the University of Chicago and is publicly available\cite{Fryxell00}. It is fully parallelized by using the Paramesh library\cite{MacNeice00}, which implements a block-structured adaptive Cartesian grid with the Message-Passing Interface (MPI) library. The FLASH architecture allows arbitrarily many alternative subprograms to coexist and to interchange with each other. The code includes multispecies processing. At the present time it includes two species: these are neutral hydrogen atoms and hydrogen ions, which seems appropriate for the study of hydrogen-dominated upper atmospheres of exoplanets.

 The application of the FLASH code to the upper planetary atmosphere physics has been described in detail in previous works\cite{H08,E10,K13}. Here we summarize the key points related to the code and describe in detail the calculation of the Ly$\alpha$ absorption.
%-----------------------------
The model includes the following processes/forces that may act on an exospheric atom:
\begin{enumerate}
   \item Collision with a UV photon, which can occur if the particle is outside of the planet's shadow. According to Hodges\cite{Hodges94}, a stellar photon can be absorbed by an atom and then consequently reradiated. This leads to an acceleration of the hydrogen atom away from the star by radiation pressure. Due to the Doppler shift of the Ly$\alpha$ frequency in the moving reference frame of an atom, the rate $\tau_{\rm rp}(v_r)$ is radial velocity dependent and proportional to the stellar Ly$\alpha$ flux at the corresponding shifted frequency. 
   \item Photoionization by a stellar photon: the reaction ${\rm H} + h \nu \rightarrow {\rm H}^+ + e$ occurs at a rate of $\tau_{\rm pi}$ when an exospheric hydrogen atom is outside the optical shadow of the planet. Then the metaparticle (see below) is removed from the simulation.
   \item Electron impact ionization, which may take place only outside the magnetosphere of the planet, where the stellar wind is present: ${\rm H} + e \rightarrow {\rm H}^+ + 2e$. The rate of the reaction $\tau_{\rm ei}$ is taken from Holzer\cite{Holzer77}. This reaction is handled numerically similar to photoionization (see below).
   \item charge exchange with a stellar wind proton, which can happen if the hydrogen atom is outside the obstacle, it can charge exchange with a stellar wind proton, producing an energetic neutral atom (ENA).
   \item Elastic collision with another hydrogen atom.
\end{enumerate}

%------------------------------------------------------------------
%\begin{figure}[htb!]
\begin{figure}[ht]
  \centering
  \includegraphics[width=0.6\textwidth]{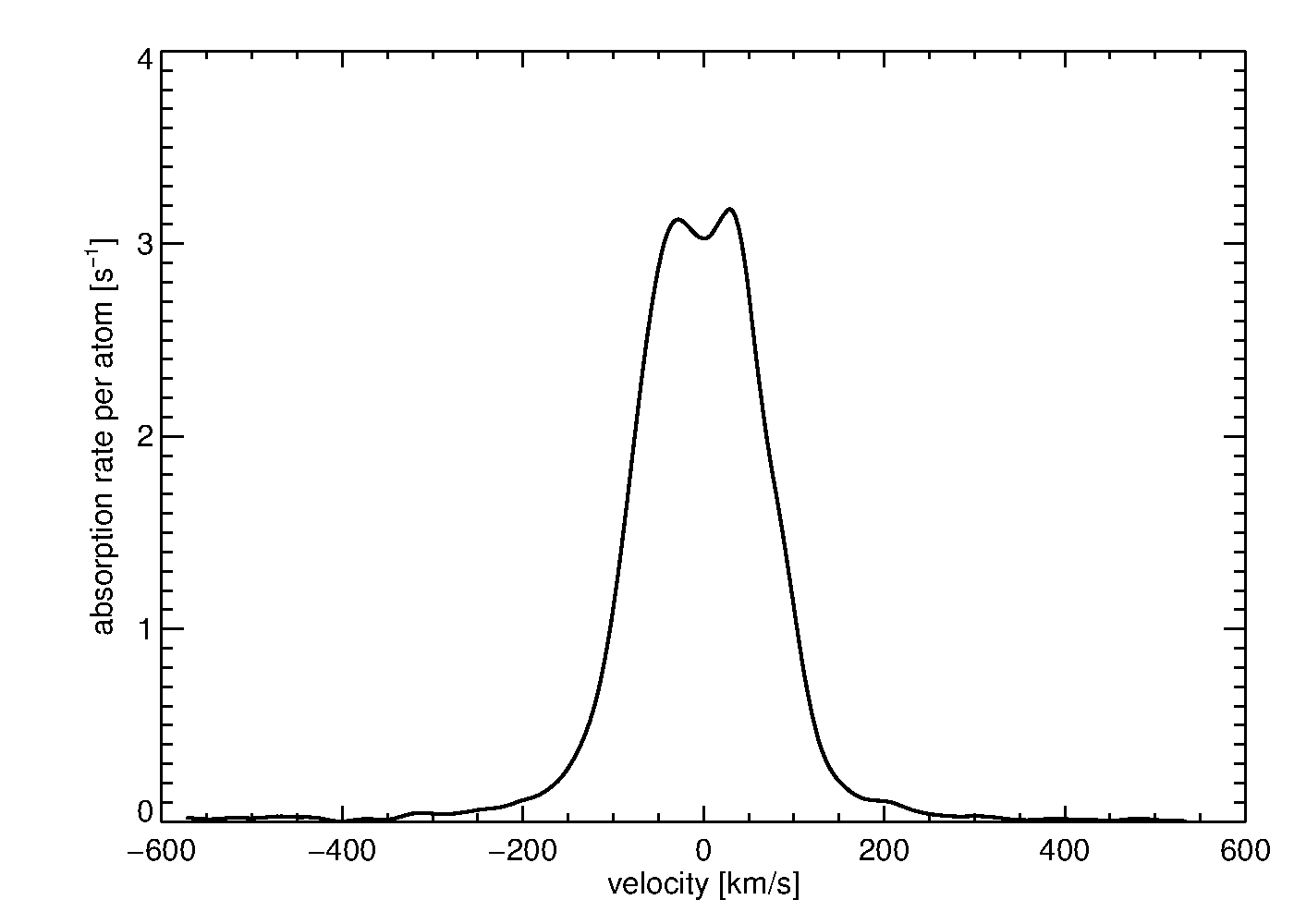}
  \caption{Ly$\alpha$ absorption rate dependent on the radial velocity of hydrogen atoms. Positive velocities denote motion towards the star according to Eq.~\ref{e_ftov} (see below). Ly$\alpha$ profile of HD 209458 was reconstructed by Brian Wood\cite{Wood05b}.}
  \label{f_rp}
\end{figure}
%-----------------------------------------------------------------

For the Ly$\alpha$ absorption and ionization rates ($\tau_{\rm i}$) after each time step, for each metaparticle, a random time is drawn from an exponential distribution with mean $\tau_{\rm i}$. The event occurs if this time is smaller than the time step. Unlike electron impact ($\tau_{\rm ei}$) and photoionization ($\tau_{\rm pi}$) rates, radiation pressure is velocity dependent and acts only on neutral hydrogen atoms with radial velocities falling inside the Ly$\alpha$ acceleration domain (Fig.~\ref{f_rp}). Depending on the radial velocity of every hydrogen neutral and the stellar Ly$\alpha$ flux, the corresponding $\tau_{\rm rp}(v_r)$ is chosen. 

For correct calculation of radiation pressure one needs a reconstructed Ly$\alpha$ profile, which represents the Ly$\alpha$ line that HD~209458b would see, without absorption by the interstellar medium and contamination by geocoronal emission. This contamination is responsible for the deep gap in a narrow band near the line center (the gap arises after subtraction of a high radiation peak produced by contaminating atoms\cite{VM03,BJ10}). Interstellar deuterium absorption must also be corrected for. The method of profile reconstruction and contamination removal has been described and successfully used for different stars\cite{Wood05}, including HD~209458\cite{Wood05b}. The photoabsorption rate is proportional to the Ly$\alpha$ flux at a particular velocity and can be easily calculated afterwards by multiplication of the 
Ly$\alpha$ spectrum with the photon absorption crossection (Fig.~\ref{f_rp}).

%All rates are scaled according to the orbital location of HD~209458b which orbits a solar-like star at a distance of $\sim$0.047~AU.

The coordinate system is centered at the center of the planet, the $x$-axis is pointing towards the center of mass of the system, the $y$-axis points in the opposite direction to the planet's orbital velocity and the $z$-axis is parallel to the direction of the angular velocity of rotation $\Omega$. $M_{\rm pl}$ and $M_{\rm st}$ are the planetary and stellar masses respectively. The outer boundary of the simulation domain is the box 
$x_{\mathrm{min}} \leq x \leq x_{\mathrm{max}}$,~ 
$y_{\mathrm{min}} \leq y \leq y_{\mathrm{max}}$, and 
$z_{\mathrm{min}} \leq z \leq z_{\mathrm{max}}$. The inner boundary is a sphere of radius $R_{\rm ib}$. 

Tidal potential, Coriolis and centrifugal forces as well as the gravitation of the star and planet acting on a neutral hydrogen atom are included following Chandrasekhar\cite{Chandrasekhar63}:

%\begin{equation}
%  \frac{d v_{x,y,z}}{dt} = \frac{\partial}{\partial x,y,z} \left[ \frac{1}{2}\Omega^2 \left(x^2+y^2 \right)+ \mu\left(x^2-\frac{1}{2}y^2 -\frac{1}{2}z^2 \right) + \left(\frac{G M_{\rm st}}{R^2} - \frac{M_{\rm st} R}{M_{\rm pl} + M_{\rm st}}\Omega^2 \right)x \right] +  2\Omega \epsilon_{(x,y,z)l3} v_l
%  \label{e_SCG}
%\end{equation}

\begin{multline}
\frac{d v_{x,y,z}}{dt} = \frac{\partial}{\partial x,y,z} \Bigg[ \frac{1}{2}\Omega^2 \left(x^2+y^2 \right)+ \mu\left(x^2-\frac{1}{2}y^2 -\frac{1}{2}z^2 \right) + \\
\left(\frac{G M_{\rm st}}{R^2} - \frac{M_{\rm st} R}{M_{\rm pl} + M_{\rm st}}\Omega^2 \right)x \Bigg] +  2\Omega \epsilon_{(x,y,z)l3} v_l
  \label{e_SCG}
\end{multline}

Here $v_{x,y,z}$ are the components of the velocity vector of the particle, $G$ is Newton's gravitational constant, $R$ is the distance between the centers of mass of the star and the planet, $\epsilon_{(x,y,z)l3}$ is the Levi-Civita symbol, and $\mu = G M_{\rm st}/R^3$. The first term on the right-hand side of the Eq.~\ref{e_SCG} represents the centrifugal force, the second is the tidal-generating potential, the third is the gravitation of the planet's host star and the planet, and the last term stands for the Coriolis force. The self-gravitational potential of a particle is neglected. Tidal force leads to the extension of the atmosphere towards and away from the host star and in extreme cases to Roche lobe overflow. 

 At the start of the simulation, the domain is empty of particles. Then neutral H atoms are launched with the rate of 600 metaparticles per second into the simulation domain from the inner boundary $R_{\rm ib}$. All particles are represented by so-called metaparticles, each corresponding to $N_m$ real particles. Protons and neutrals always have the same weights, but $N_m$ can change for different inner boundary and stellar wind densities, i.e., it differs from simulation to simulation. The local hydrogen density determines the probability that a particle is launched at a specific location on the inner boundary. The velocity of a particle is also random and taken from the probability distribution proportional to $(\mathbf{n} \times \mathbf{v})\exp(-a|\mathbf{v}|^2)$, where $\mathbf{n}$ is the surface normal, $\mathbf{v}$ is the velocity of the particle, and $a = m/2kT_{\rm ib}$. Here $m$ is the mass of a neutral H atom, $k$ is Boltzmann's constant, and $T_{\rm ib}$ is the temperature at the inner boundary. Although the used distribution is not a Maxwellian, the flux through a surface gives a Maxwellian distribution at the location\cite{Garcia00} with the particle flux through the surface of $n_{\rm ib}/\sqrt{\pi a}$ and a total production rate\cite{H08} of $n_{\rm ib} R_{\rm ib}^2 \sqrt{8 \pi k T_{\rm ib}/m}$~s$^{-1}$. Here $n_{\rm ib}$ denotes the inner boundary density.

The protons' velocities are obtained from a Maxwellian distribution with the temperature $T_{\rm sw}$ and bulk velocity $v_{\rm sw}$. A uniform stellar wind eventually builds up in the $x$-``shadow'' cells just outside the simulation domain, and is then moved into the domain. The relative velocity at the planet, $\mathbf{v}_{\rm rel}$, is related to the stellar wind velocity and the planet's orbital velocity by $\mathbf{v}_{\rm rel} = \mathbf{v}_{\rm sw} +\mathbf{v}_{\rm pl}$. The protons cannot penetrate inside a planetary magnetosphere. When they reach the magnetosphere boundary, they can be reflected or rejected from the simulation domain.

 Trajectories of all particles are numerically integrated with the time step of 25 seconds according to all forces acting on an atom or a proton. The time advance of the particles from time $t$ to time $t+\Delta t$, is done using the symplectic integrators derived by Candy and Rozmus\cite{Candy91}, 
\begin{equation}\label{e_CR}
  \begin{array}{ccl}
    \mathbf{x} & \leftarrow & \mathbf{x} + c_k \Delta t \mathbf{v}, \\
    \mathbf{a} & \leftarrow & \mathbf{a}(\mathbf{x},t), \\
    \mathbf{v} & \leftarrow & \mathbf{v} + d_k \Delta t \mathbf{a}, \\
    t          & \leftarrow & t + c_k \Delta t, 
  \end{array}
\end{equation}
for $k=1,\ldots n$. Here $\mathbf{x}$ are the particle positions, $\mathbf{v}$ the velocities, and $\mathbf{a}(\mathbf{x},t)$ the accelerations. 
The coefficients $c_k$ and $d_k$ can be found in Candy and Rozmus\cite{Candy91}. %are listed in Table~\ref{tab:CR} for $n=2$ and 4. 
The global order of accuracy is $n$, and $n=2$ corresponds to the Leapfrog method.  In this work we have used $n=4$. 

After each time step, the positions of the particles are checked, and the particles located in the same cells undergo hard sphere collisions. All collisions are modelled by using the DSMC method. The applied collision crossections for protons and hydrogen atoms\cite{Lindsay05} $\sigma_{H^+-H}$ and for hydrogen atoms\cite{Izmodenov00} $\sigma_{H-H}$ are $2 \times 10^{-19}$~m$^2$ and $10^{-21}$~m$^2$, respectively.

The total collision frequency $\nu_c$ in a volume is\cite{Bird76,H08}
\begin{equation}
  \label{e_col}
  \frac{1}{2}n \nu_c = \frac{1}{2} n^2 \overline{\sigma v_r}.
\end{equation}
In Eq.~\ref{e_col} $n$ is the total number density of all species, $\sigma$ is the total collision crossection, and $v_r$ is the relative velocity between the particles. The bar stands for average. For each pair of particles the collisional probability is proportional to $\sigma v_r$. For each cell, $n$ is estimated by $N_p N_m /V_c$, where $N_p$ is the number of particles in the cell and $V_c$ is the cell volume.

To reduce the calculation time, which is proportional to $N_p^2$, the averages are not directly computed. Instead a maximum value $(\sigma v_r )_{\rm max}$ is estimated and used in Eq.~\ref{e_col} to compute the number of trials\cite{Garcia00}. For each trial a random pair is taken, and a random number $N$, in the interval $[0, (\sigma v_r )_{\rm max}]$, is computed. If $(\sigma v_r )_{\rm max} > N$ for the chosen pair, the collision is accepted.
The random pair above is uniformly distributed if $N_m$ is the same for all particles, which is always fulfilled in all performed simulations.

charge exchange between stellar wind protons and exospheric hydrogen atoms takes place outside a conic shaped obstacle that represents the magneto-ionopause of the planet:
\begin{equation}
  \label{e_obs}
  x = R_{\rm s} \left(1-\frac{y^2 + z^2}{R_{\rm t}^2} \right)
\end{equation}
Here $R_{\rm s}$ stands for the magnetosphere or planetary obstacle stand-off distance and $R_{\rm t}$ for the width of the obstacle. Since the obstacle shape and location depend strongly on the planetary magnetic field strength, one may model the interaction of the stellar wind with magnetized as well as with non- or weakly magnetized planets by the appropriate choice of $R_{\rm s}$ and $R_{\rm t}$.\newline
The obstacle is rotated by an angle of $\arctan (v_{\rm pl}/v_{\rm sw})$, to account for the finite stellar wind speed relative to the planet's orbital speed.
%----------------------------------
\subsection*{Post-Processing And Ly$\alpha$ Attenuation}

After the hydrogen corona around HD~209458b is simulated, we calculate the Ly$\alpha$ in-transit attenuation. At this stage one has to take into account spectral line broadening.

Real spectral lines are never absolutely ``sharp''. This is related to several broadening mechanisms:
\begin{enumerate}
  \item natural broadening;
  \item collisional broadening;
  \item Doppler or thermal broadening.
\end{enumerate}

The ``natural line width'' is a result of quantum effects and arises due to the finite lifetime of an atom in a definite energy state. A photon emitted in a transition from this level to the ground state will have a range of possible frequencies: $\Delta f \sim \Delta E / \hbar \sim 1 / \Delta t $. The distribution of frequencies can be approximated by a Lorentzian profile. %The symmetry of the Lorentz function around zero implies that if the original velocity distribution is symmetric, the broadened distribution will also be symmetric.

Collisional broadening is caused by the collisions randomizing the phase of the emitted radiation. This effect can become very important in a dense environment, yet above the exobase it does not play a role and is important only below $R_{\rm ib}$.

The third type of broadening, which plays a significant role in the upper atmosphere of a ``Hot Jupiter'', is thermal broadening. If $f_0$ is the centroid frequency of the absorption line, the frequency will be shifted due to the Doppler effect. Combining Doppler shift with the Maxwellian distribution of $v_x$, one can obtain a Gaussian profile function, which is decreasing very rapidly away from the line center.

The combination of thermal and natural (or collisional) broadening is described by the Voigt profile, which is the convolution of the Lorentz and Doppler profiles. 

%The application to the computation of Ly$\alpha$ transmissivity of a hydrogen cloud around HD 209458b in the Ly$\alpha$ line during the transit. After the computation the modelled and observed spectra may be compared. Unlike the main code, post-processing software is written in Python Programming Language, which is also a free source (\cite{python_{\rm s}ite}).

In the present study an analytical solution for the absorption profile cannot be obtained, since it is not only thermal atoms that contribute to the broadening. The presence of a non-thermal population of hot atoms (ENAs and atoms accelerated by the radiation pressure) changes the picture. Mathematically it means that the line width cannot be described by the Voigt profile anymore. We calculate the natural broadening for all atoms and bin it by velocity, which automatically gives us the Doppler broadening for a particular velocity distribution.

After a hydrogen cloud is simulated and by knowing the positions and velocities of all the hydrogen metaparticles at a certain time, we compute how these atoms attenuate the stellar Ly$\alpha$ radiation by using a post-processing software written in Python programming language. The velocity spectrum can then be converted in a frequency via the relation 
\begin{equation}
	f = f_0 + v_x / \lambda_0
	\label{e_ftov}
\end{equation}
with $f_0 = c / \lambda_0$, $\lambda_0 = 1215.65 \times 10^{-10}$~m.  To compute the transmissivity along the line-of-sight (LOS) we follow the approach of Semelin\cite{Semelin07}, where the relation between the observed intensity $I$ and the source intensity $I_0$ as a function of frequency $f$ can be written
as
\begin{equation}
	T=I/I_0=e^{-\tau(f)}=e^{-\sigma(f)Q}.
	\label{e_T}
\end{equation}

Here $\tau=\sigma(f)Q$ is the frequency dependent optical depth, $Q$ is the column number density of hydrogen atoms and $\sigma(f)$ is the frequency dependent crossection, which depends on the normalized velocity spectrum, the Ly$\alpha$ resonance wavelength and the natural absorption crossection in the rest frame of the scattered hydrogen atom\cite{Peebles93}. The quantity $T$ is called the transmissivity.

%------------------------------------------------------------------
\begin{figure}[t]
  \centering
  \includegraphics[width=0.7\textwidth]{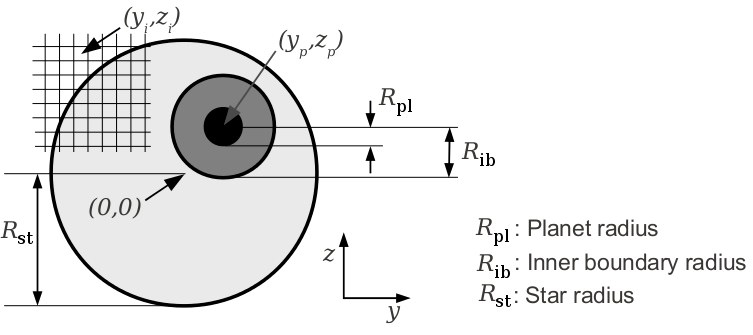}
  \caption{Computation of the Ly$\alpha$ transmissivity for the simulation domain. For each ``pixel'' (arrow marked by $y_i,z_i$) the attenuation as a function of velocity (wavelength) is computed. The attenuation equals unity for pixels covering the planet.}
  \label{f_G_Lya}
\end{figure}
%------------------------------------------------------------------

If we have now a hydrogen cloud in front of a star, the transmissivity of the stellar spectrum can be computed as illustrated in Fig.~\ref{f_G_Lya} where the $yz$-plane is discretized into a grid with $N_c$ cells. For each cell in the grid along lines of sight in front of the star ($y^2 + z^2 < R_{\rm st}^2$), the velocity spectrum of all hydrogen atoms in the column along the $x$-axis can be calculated. We assume the observer is at an infinite distance, and that the emission is uniform over the stellar disc, as seen by the observer. Then the transmissivity can be averaged over all columns in the $yz$-grid except those particles which fall outside the projected limb of the star or inside the planetary disc. The average transmissivity is then calculated as 
\begin{equation}
	\bar{T}(f) = \frac{1}{N_c} \displaystyle\sum_{i=1}^{N_c} T_i(f).
	\label{e_avT}
\end{equation}

For lines-of-sight in front of the planet $(y - y_p )^2 + (z - z_p )^2 < R_{\rm pl}^2$, where ($y_p$ , $z_p$) is the planet center position, we set $T_i = 0$ (zero transmissivity). The average transmissivity is then applied to the observed out-of-transit spectrum $I_0$ yielding the modelled in-transit spectrum
\begin{equation}
	I_m (f ) = I_0 (f )\bar{T}(f).
	\label{e_mSp}
\end{equation}

The frequency dependent crossection is defined by:
\begin{equation}
		\sigma(f)=\int_{-\infty}^{+\infty}\check{u}(v_x)~\sigma_N[(1+v_x/c)f]dv~[m^2],
\end{equation}
where $\check{u}(v_x)$ is the normalized velocity spectrum along the LOS, so that $\int_{-\infty}^{+\infty}\check{u}(v_x)dv_x=1$, $c$ - the speed of light.

The natural absorption crossection in the rest frame of the scattered hydrogen atom is taken according to Peebles\cite{Peebles93}
\begin{equation}
	\sigma_N(f)=\frac{3 \lambda_{\alpha}^2 A^2_{21}}{8\pi} \frac{(f/f_{\alpha})^4}{4\pi^2(f-f_{\alpha})^2+(A_{21}^2/4)(f/f_{\alpha})^6}~[m^2]
	\label{e_crossec}
\end{equation}
Here $f_{\alpha}=c/\lambda_{\alpha}$ with $\lambda_{\alpha}=1.21576 \times 10^{-7}$~m the Ly$\alpha$ resonant wavelength and $A_{21}=6.265 \times 10^8$~s$^{-1}$ the rate of radiative decay from the $2p$ to the $1s$ energy level.

Multiplying $\sigma(f)$ with Q, the optical depth is computed directly without normalizing the velocity spectrum:
\begin{equation}
	\tau(f)=\int_{-\infty}^{+\infty} u(v_x)~\sigma_N[(1+v_x/c)f]dv.
	\label{e_tau}
\end{equation}

The expression for the natural absorption crossection given in Eq.~\ref{e_crossec} is approximated by the Lorentzian profile
\begin{equation}
	\sigma_N(f)=f_{12} \frac{\pi e^2}{m_e c} \frac{\Delta f_L/2\pi}{(f-f_{\alpha})^2 + (\Delta f_L /2)^2},
	\label{e_crossLor}
\end{equation}
where $f_{12}= 0.4162$ is the Ly$\alpha$ oscillator strength, $e$ is the elementary charge, $m_e$ is the electron mass, and $\Delta f_L = 9.936 \times 10^7$~[s$^{-1}$] is the natural line width\cite{Semelin07}. We used the crossection from Eq.~\ref{e_crossLor} to include the contribution of broadening to the absorption.

If we disregard natural broadening, we can approximate the crossection by a delta function. Then the optical depth is directly proportional to the velocity spectrum, and can be approximated by
\begin{equation}
	\tau(f) \approx Cu[\lambda_\alpha(f_\alpha -f)],~\mathrm{where}~C= \int_{-\infty}^{\infty} \sigma_N[(1 + v_x/c)f]dv_x.
	\label{e_crossDelta}
\end{equation}
We used Eq.~\ref{e_crossDelta} to compute spectra without broadening to study the absorption caused by the atmosphere and ENAs alone.

To account for the contribution of the lower atmosphere, a Maxwellian velocity spectrum corresponding to a hydrogen gas with a specified column density and temperature, is added to all pixels inside the inner boundary $R_{\rm ib}$. The atmospheric spectrum $u_{\rm atm} (v_x)$ is added to $u(v_x)$ in Eq.~\ref{e_tau}. We assume an isothermal atmosphere with density and temperature given at $R_{\rm ib}$. Then the density as a function of as a function of radial distance from the center of the planet, $r$, inside $R_{\rm ib}$ is given by

\begin{equation}
	n(r) = n (R_{\rm ib}) \exp \left(- (r - R_{\rm ib})/H \right),~~H = \frac{kT_{\rm ib}}{mg}.
	\label{e_plsp}
\end{equation}
Here $H$ is the scale height. The gravitational acceleration $g$ is computed at $R_{\rm ib}$ as $g = GM_{\rm pl} /R_{\rm ib}^2$. The probability function for one velocity component is taken from a Maxwellian distribution. The column density $q$ of the atmosphere is computed by numerical integration of Eq.~\ref{e_plsp} along the LOS inside $R_{\rm ib}$. The probability function for one velocity component for a Maxwellian gas is\cite{Semelin07}
\begin{equation}
	p(v_x) = \frac{1}{\sqrt \pi v_t} \exp (-v_x^2/v_t^2),~\mathrm{with}~v_t=\sqrt{2kT_{\rm ib}/m}
	\label{e_velsp}
\end{equation}

The contribution of the atmosphere to the spectrum is then $u_{\rm atm} (v_x) = qp(v_x) $.

%----------------------------------
\subsection*{Numerical Issues}

%One of the interesting numerical problems, which was solved until obtaining results, consisted in insufficient computing power available for modelling of the hydrogen clouds around HD 209458b.
One of the numerical challenges we had when modelling the hydrogen cloud around HD 209458b was the large amount of computing power required.

In the above computations the data structure used is a three dimensional array corresponding to the $(v, y, z)$ coordinates, with cell sizes $\Delta v$, $\Delta y$, and $\Delta z$. Since we are averaging the exponent of the optical depth (Eq.~\ref{e_T}), the sampling is very important and all weight in one bin is very different from having it spread out, so that absorption computed for two particles with half-weight in neighbouring cells does not equal absorption computed for only one particle with full weight in one of the cells. Insufficient computing power decreases the maximum number of metaparticles, increases their weight and prevents the presence of metaparticles in all velocity bins and spatial cells.

If there are few ENA metaparticles, one has to have a large spatial cell size (pixel size) and large velocity bin size. For each velocity bin the cloud of particles should be continuous. Gaps in the cloud lead to bad statistics and unphysical effects (for example, less transmissivity for higher densities). One has to plot the cloud for each velocity bin to make sure that the statistics are good enough (i.e., the cloud is continuous).

Essentially, in the cloud, each $u(v, y, z)$ bin has to have particles in it to get good statistics. The discrete approximation (Eq.~\ref{e_avT}) of the average transmissivity will approach the correct solution when the cell size approaches zero and the number of particles approaches infinity. This represents a tradeoff between the errors from the finite cell size and the number of particles. Given a situation with continuous clouds of particles in each velocity bin, one has to increase the number of particles by a factor of 8 if the cell size ($\Delta v$, $\Delta y$, and $\Delta z$) is reduced by a factor of two. One way of increasing the number of particles is to average the solution over many time steps. The distance in time between the solutions used should then be large enough to avoid too much correlation between the solutions. Another way to improve the statistics is to use a more accurate deposit of the particles onto the $(v, y, z)$ grid. At the moment, particles are deposited in the cell according to the Cloud in cell (CIC) algorithm, where each particle is viewed as a cloud of particles with the same size as the cell. We only apply CIC for the spatial coordinates, $(y, z)$. In practice each particle weight will then be distributed to four cells. This is a self consistent smoothing of the solution and should improve statistics. Even higher order deposits should be possible.

Another numerical issue arises from the integration of Eq.~\ref{e_tau}, since the natural absorption crossection, $\sigma_N(f)$, is a very peaked function, approaching a delta function. This requires a very fine velocity grid when numerically integrating Eq.~\ref{e_tau}. Using the cell width, $\Delta v$, as the step in the integration gives a too coarse grid, since $\Delta v$ has to be large enough to get good statistics, as explained above. The solution is to upsample $u(v)$ in Eq.~\ref{e_tau} before doing the numerical integration. In the present study we used upsampling before computing the Ly$\alpha$ absorption (usually by a factor not less that 400) and downsampling afterwards.

Resampling algorithms are used in signal processing and usually denote the change of the sampling rate of the signal. The upsampling factor $n$ is usually an integer or a rational fraction greater than unity. The sampling rate is multiplied by this factor. In the present case, it was used for resampling on the velocity grid. By upsampling, $n-1$ additional points are inserted into an array of values, afterwards the interpolation is fulfilled. This increases the number of metaparticles for processing by $n$. Downsampling is performed after the Ly$\alpha$ absorption is computed to make the further processing of the array easier.

%----------------------------------
\subsection*{Simulation Parameters}

%STAR PARAMETERS
HD~209458 is a G0V Sun-type dwarf star with an age comparable with that of the Sun (4$\pm$2~Gyr). Since many parameters of HD~209458 are unknown, this similarity allows one to scale the known solar values from 1~AU to the orbital location of HD~209458b at 0.047~AU (e.g., photo- and electron impact ionization rates). We have therefore used the solar wind parameters for the estimation of HD~209458's stellar wind density, temperature and velocity. Since HD~209458 is a Sun-type star, it is possible that its wind, radiation parameters and magnetic field strength change within an activity cycle similar to the solar cycle. However, there is little known about the existence, amplitude and period of cycles on other solar-like stars. The measurements performed for some tens of targets show that approximately 61\% of the observed stars were cycling\cite{Lovis11}. It would also be very difficult to constrain the phase of the cycle when the observations were made. Because of this, the mean solar wind values for 0.047~AU with slightly increased stellar wind velocity were used\cite{Priest82}. We increased the stellar wind velocity since as we show below, the observation cannot be fitted with slower stellar wind. However, the assumed value is still in the range for the slow solar wind\cite{Priest82}. For comparison, the standard upstream parameters can also be found in \cite{Baumjohann96}.
%------------------------------------------------------------------
\begin{figure}[t]
  \centering
  \includegraphics[width=0.4\textwidth]{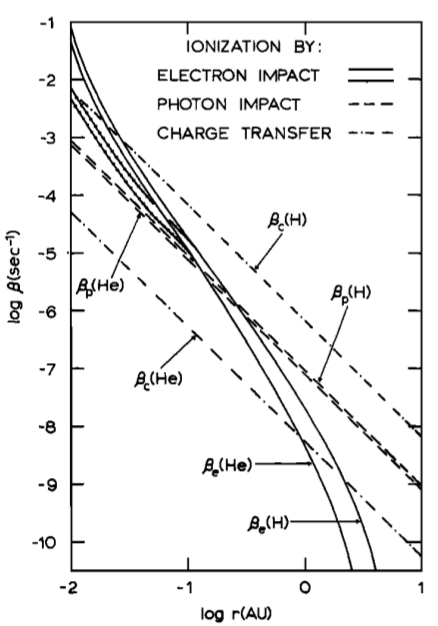}
  \caption{Ionization rate $\beta$(s$^{-1}$) (defined in the present paper as $\tau$) as a function of heliocentric radial distance $r$(AU) for ionization of H and He by electron impact, photon impact and charge transfer\cite{Holzer77}. Electron impact ionization rates (defined in the present paper as $\tau_{\rm ei}$) are shown for the temperature profile and both density profiles given in Fig.~1b\cite{Holzer77} (see Fig.~1a\cite{Holzer77}).}
  \label{f_Holzer}
\end{figure}
%------------------------------------------------------------------

For a Sun-type star, electron impact ionization plays a significant role only for planets located very close to their host stars\cite{Holzer77}. However, this type of ionization was disregarded in all previous studies for HD~209458b. In this study, we used the solar values scaled to the appropriate distance from the star. Fig.~\ref{f_Holzer} illustrates the dependence of various ionization rates in the vicinity of the Sun on the distance. We have adopted an electron impact ionization rate $\tau_{\rm ei}$ at 0.047~AU of $\sim 1.25 \times 10^{-4}$~s$^{-1}$ according to the chosen stellar wind density of $5 \times 10^9$~m$^{-3}$. The photon ionization rate $\tau_{\rm pi}$ 
at 0.047~AU was chosen the same way\cite{Holzer77} and is equal to $\sim 6.0 \times 10^{-5}$~s$^{-1}$.

%%------------------------------------------------------------------
%\begin{figure}[t]
%  \centering
%  \includegraphics[width=0.6\textwidth]{HD209458b_atm}
%  \caption{Structure of the atmosphere of the "Hot Jupiter" HD 209458b.}
%  \label{f_HDatm}
%\end{figure}
%%------------------------------------------------------------------

To model the interaction between the upper atmosphere of an exoplanet and the stellar wind one obviously needs some input parameters for the planetary atmosphere. In the current study, the computational domain includes the upper layers of the atmosphere. We start at the level of $2.8~R_{\rm pl}$. At these altitudes we assume the atmospheric composition is mainly atomic hydrogen, although the other species are also present in small amounts\cite{Linsky10,Koskinen13}. We took the hydrogen density and temperature at the inner boundary level from the atmosphere model\cite{Koskinen13}, which is a sophisticated photochemical-hydrodynamical model of the thermosphere of HD~209458b. This model solves the set of hydrodynamic equations for a XUV heated and expanded upper atmosphere composed of several neutral and ionized species. The results of this model can be considered accurate only in the regions where Knudsen number $Kn = \Lambda/h < 1$, where $\Lambda$ is the mean free path and $h$ is the local density scale height, i.e. in the collision-dominated atmosphere. This model also confirms that molecules dissociate near the 1 $\mu$bar level, so that there is no need to include complex molecular chemistry above this level, and also predicts that a significant fraction of the hydrogen remains neutral at least up to 3$R_{\rm pl}$ level, which is in agreement with our assumptions made for the exosphere in the present work. The model\cite{Koskinen13} predicts an upper limit of $8 \times 10^3$~K for the mean (pressure averaged) temperature below 3$R_{\rm pl}$, with a typical value of $7 \times 10^3$~K and a lower limit of $6 \times 10^3$~K based on the average solar XUV flux at 0.047~AU. For that reason, the inner boundary temperatures used in the simulations were fixed in this range. Inner boundary hydrogen density was varied in the range of $\sim$1.0--4.0$\times 10^{13}$~m$^{-3}$. In our best fit, we set the inner boundary temperature to $6 \times 10^3$~K and the inner boundary density to $2.0 \times 10^{13}$~m$^{-3}$ to avoid overabsorption due to spectral broadening (see description below), but temperatures up to $8 \times 10^3$~K are not excluded.

\begin{figure}[t]
  \centering
  \includegraphics[width=1.0\textwidth]{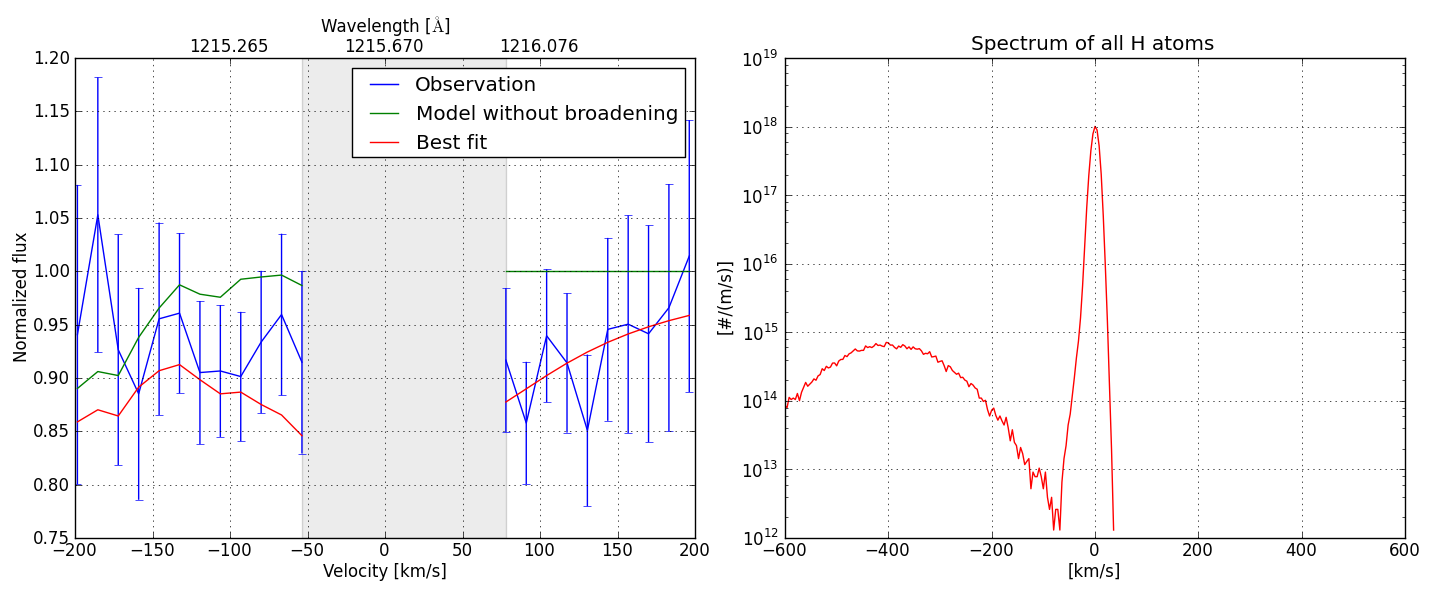}
  \caption{Left panel: comparison of modelled and observed Ly$\alpha$ spectra at mid-transit. The region of geocoronal emission at low velocities is excluded because of contamination. The velocity spectrum is converted in a frequency via Eq.~\ref{e_ftov}. Blue and red lines here and in all figures below: blue, the observed profile during the transit, normalized by the out-of-transit profile. Red, the modelled profile, computed including charge exchange, radiation pressure with self-shielding, photo- and electron impact ionization, and natural broadening (best fit, $\chi^2 \approx 0.081$). Green, the same as red, but excluding broadening ($\chi^2 \approx 0.139$). The abscissa in all figures is the hydrogen velocity along the $x$ axis (positive away from Earth, towards the star). One can clearly see that natural broadening contributes on both wings of Ly$\alpha$ line, while other effects contribute to the absorption on the blue wing produced by atoms moving away from the star. Right panel: the corresponding velocity spectrum along the $x$ axis. Here and in all figures below the velocity spectrum of the best fit is plotted with the red line.}
  \label{f_broad_Lya}
\end{figure}

The physical parameters and values used in the simulation which showed the best agreement with the observations (best fit parameters) are presented in Table~\ref{t_mod1}. Numerical parameters related to the code are presented in Table~\ref{t_mod2}. These parameters were used also in the other simulations unless otherwise noted. The corresponding calculated Ly$\alpha$ in-transit spectrum with and without broadening in comparison to the observation is shown in Fig.~\ref{f_broad_Lya}.

%---------------------------------------------------------------------
\begin{table}
\begin{minipage}{\textwidth}
  \caption[]{Default values of physical parameters (best fit), and values of constants used in the simulations, unless otherwise noted. }
  $$ 
  \begin{array}{p{0.5\linewidth}lll}
    \hline
    \noalign{\smallskip}
    Name &  \mathrm{Symbol} & \mathrm{Value} \\
    \noalign{\smallskip}
    \hline
    \noalign{\smallskip}
Star mass		 	& M_{\rm st} 	& 2.28\times 10^{30}~\mathrm{kg} & \approx 1.148~M_{\rm Sun}\\
Planet radius 		 	& R_{\rm pl}	& 9.54\times 10^7~\mathrm{m} & \approx 0.71~M_{\rm Jup} \\
Planet mass 		 	& M_{\rm pl}	& 1.21\times 10^{27}~\mathrm{kg} &\approx 1.38~R_{\rm Jup}\\
Orbital distance	 	& 		& 7.1\times 10^9~\mathrm{m} & \approx 0.047~\mathrm{AU}\\
Angular velocity 	 	& \Omega	& 2\times 10^{-5}~\mathrm{rad/s} & \\
Inner boundary radius    	& R_{\rm ib}	& 2.7\times 10^8~\mathrm{m} & \approx 2.8~R_{\rm pl} \\
Inner boundary temperature	& T_{\rm ib}	& 6\times 10^3~\mathrm{K} & \\
Inner boundary density 		& n_{\rm ib}		& 2\times 10^{13}~\mathrm{m}^{-3}& \\
Obstacle standoff distance	& R_{\rm s} 	 	      & 2.76 \times 10^8 \mathrm{m} & \approx 2.9~R_{\rm pl} \\
Obstacle width			& R_{\rm t} 	  	      & 2.86 \times 10^8 \mathrm{m} & \approx 3.0~R_{\rm pl} \\
H-H crossection 		&		& 10^{-21}~\mathrm{m}^2 & \\
H-H$^+$ crossection\footnote{Energy dependent\cite{Lindsay05}. The appropriate value for considered energies of protons is chosen.} &      & 2\cdot 10^{-19}~\mathrm{m}^2 & \\
UV absorption rate 		& \tau_{\mathrm{rp}}  &  \mathrm{velocity~dependent} & {\rm see~Fig.~\ref{f_rp}}\\
Photoionization rate 		& \tau_{\mathrm{pi}}  & 6.0\times 10^{-5}~\mathrm{s}^{-1} & \\
Electron impact ionization rate	& \tau_{\mathrm{ei}}  & 1.25\times 10^{-4}~\mathrm{s}^{-1} & \\
Stellar wind density		& n_{\mathrm{\rm sw}}     & 5\times 10^9~\mathrm{m}^{-3} & \\
Stellar wind velocity		& u_{\mathrm{\rm sw}}     & 400 \times 10^3~\mathrm{m/s} & \\
Stellar wind temperature		& T_{\mathrm{\rm sw}}     & 1.1 \times 10^6~\mathrm{K} & \\
    \noalign{\smallskip}
    \hline
  \end{array}
  $$ 
\label{t_mod1}
\end{minipage}
\end{table}
%---------------------------------------------------------------------

%---------------------------------------------------------------------
\begin{table}
  \caption[]{Default numerical parameter values used in the simulations (best fit), 
             unless otherwise noted. }
  $$ 
  \begin{array}{p{0.5\linewidth}lr}
    \hline
    \noalign{\smallskip}
    Name &  \mathrm{Symbol} & \mathrm{Value} \\
    \noalign{\smallskip}
    \hline
    \noalign{\smallskip}
 & x_{\mathrm{min}} & -6 \times 10^9 \mathrm{m} \\
 & x_{\mathrm{max}} &  2 \times 10^9 \mathrm{m} \\
 & y_{\mathrm{min}} & -5 \times 10^9 \mathrm{m} \\
 & y_{\mathrm{max}} &  5 \times 10^9 \mathrm{m} \\
 & z_{\mathrm{min}} & -3.5 \times 10^9 \mathrm{m} \\
 & z_{\mathrm{max}} &  3.5 \times 10^9 \mathrm{m} \\
Number of particles per metaparticle (best fit) & N_{\rm m} & 2.58132 \times 10^{32} \\
Number of cells (best fit) & N_{\rm c} & 1.825 \times 10^7 \\
Final time & t_{\mathrm{max}} & 10^5 \mathrm{s} \\
Time step & \Delta t & 25~\mathrm{s}\\
    \noalign{\smallskip}
    \hline
  \end{array}
  $$ 
\label{t_mod2}
\end{table}
%---------------------------------------------------------------------

\begin{figure}[t]
  \centering
  \includegraphics[width=0.95\textwidth]{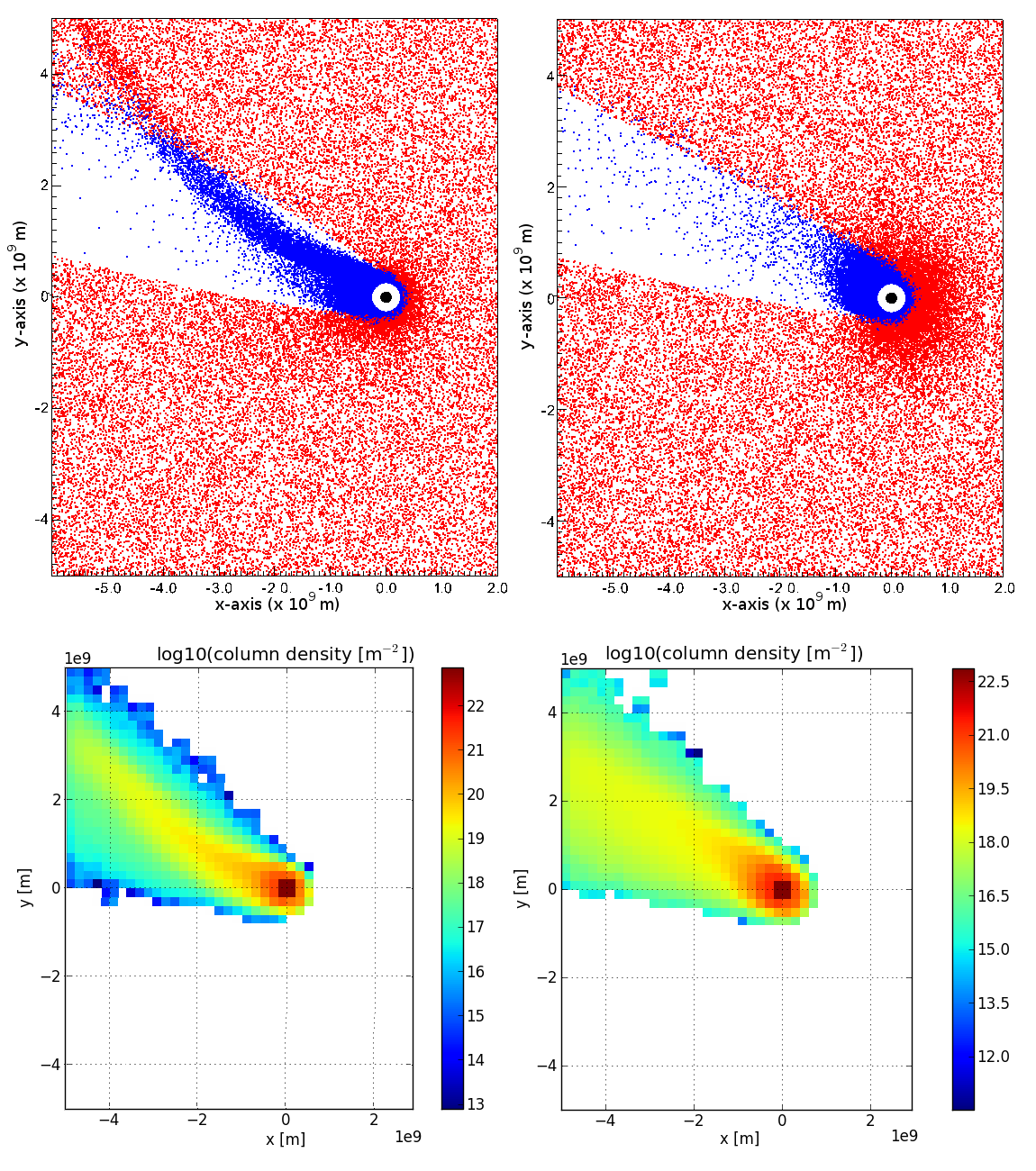}
  \caption{Top: comparison of modelled neutral hydrogen clouds around HD~209458b with (right panel) and without (left panel) self-shielding. Presented are slices of the modelled 3D atomic corona around the planet for $-10^7 \le z \le 10^7$~m. Blue and red dots correspond to neutral hydrogen atoms and hydrogen ions, which include stellar wind protons, respectively. The black dot in the center represents the planet. The white empty area around the planet corresponds to the XUV heated thermosphere up to the height $R_{\rm ib}$. Bottom: the corresponding column density with  (right panel) and without (left panel) self-shielding.}
  \label{f_shield}
\end{figure}

%SELF-SHIELDING: DESCRIBE HERE
In case of HD~209458b, the self-shielding effect significantly diminishes the acceleration by radiation pressure and photoionization of neutral hydrogen atoms\cite{BL13}, and decreases the length of the accelerated neutral hydrogen tail behind the planet. Self-shielding means that the medium becomes optically thick in Ly$\alpha$, which prevents the UV photons from penetrating inside the deeper layers and interacting with hydrogen atoms. In the present study, we do not include either UV scattering or photoionization in the atmosphere below $R_{\rm ib}$, which is protected from the UV radiation by the upper atmosphere. However, self-shielding is also important in the exosphere where we introduce a corresponding exponential weakening factor. This factor is defined by the optical depth towards the star. We estimated the value of the self-shielding parameter from a run with full radiation pressure acceleration. According to this estimate, the neutral hydrogen tail of HD~209458b stays optically thick in Ly$\alpha$ up to the distance of 20~$R_{\rm pl}$ (Fig.~\ref{f_shield}). For this reason we chose this value as self-shielding parameter in our best fit simulation. Inside this sphere, the probability for a UV photon to be scattered on an exospheric neutral is exponentially decreased and equals zero at the level of $R_{\rm ib}$.

%------------------------------------------------------------------
%------------------------------------------------------------------

%------------------------------------------------------------------
%------------------------------------------------------------------

\begin{figure}[t]
  \centering
  \includegraphics[width=1.0\textwidth]{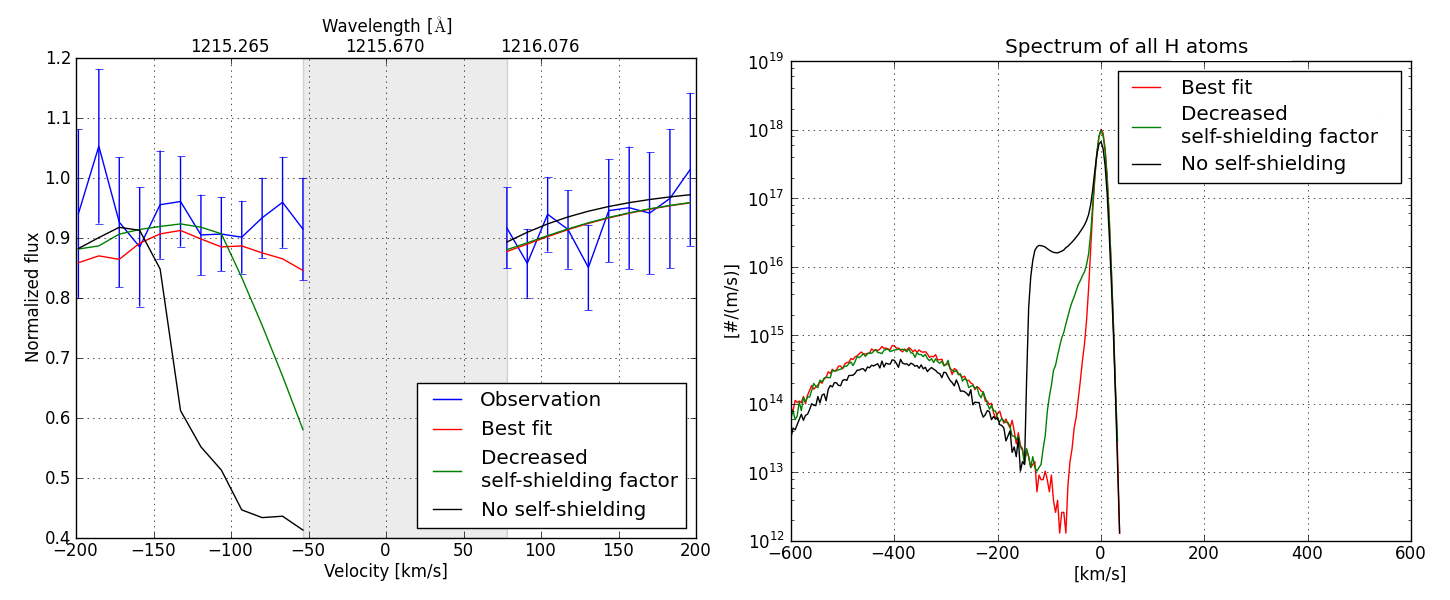}
  \caption{Left panel: the green line shows the best fit model (see the caption of Fig.~\ref{f_broad_Lya}), but with a smaller self-shielding parameter equal to $10 R_{\rm pl}$ ($\chi^2 \approx 0.300$). The black line shows the case of no self-shielding ($\chi^2 \approx 1.544$). Right panel: the corresponding velocity spectra along the LOS. As can be seen without self-shielding the bulk atmosphere undergoes acceleration by the stellar UV radiation.}
  \label{f_shield_Lya}
\end{figure}

For comparison, we show also a spectrum calculated with the same parameters, but with a weaker self-shielding factor of only 10~$R_{\rm pl}$ (Fig.~\ref{f_shield_Lya}). As can be clearly seen, a smaller self-shielding factor (i.e., stronger acceleration by the radiation pressure) leads to strong overabsorption and does not allow us to reproduce the observation. The velocity spectrum shows a wider main peak shifted towards negative velocities.

%------------------------------------------------------------------
%------------------------------------------------------------------

%----------------------------------
\subsection*{Influence of Changing the Parameters}

In this section, we discuss the influence of changing the parameters and calculate for each simulation, the chi-squared goodness of fit as $\chi^2 = \sum\limits_{i} \frac{(M_i - O_i)^2}{O_i}$, where $O_i$ is the observed value of one particular point, and $M_i$ is the calculated value at the same point of the modelled absorption curve. The summation is made over the number of analysed data points in the observation. The observational error is not taken into account. Our best fit has a $\chi^2$ of 0.081.

%------------------------------------------------------------------
\begin{figure}[t]
  \centering
  \includegraphics[width=1.0\textwidth]{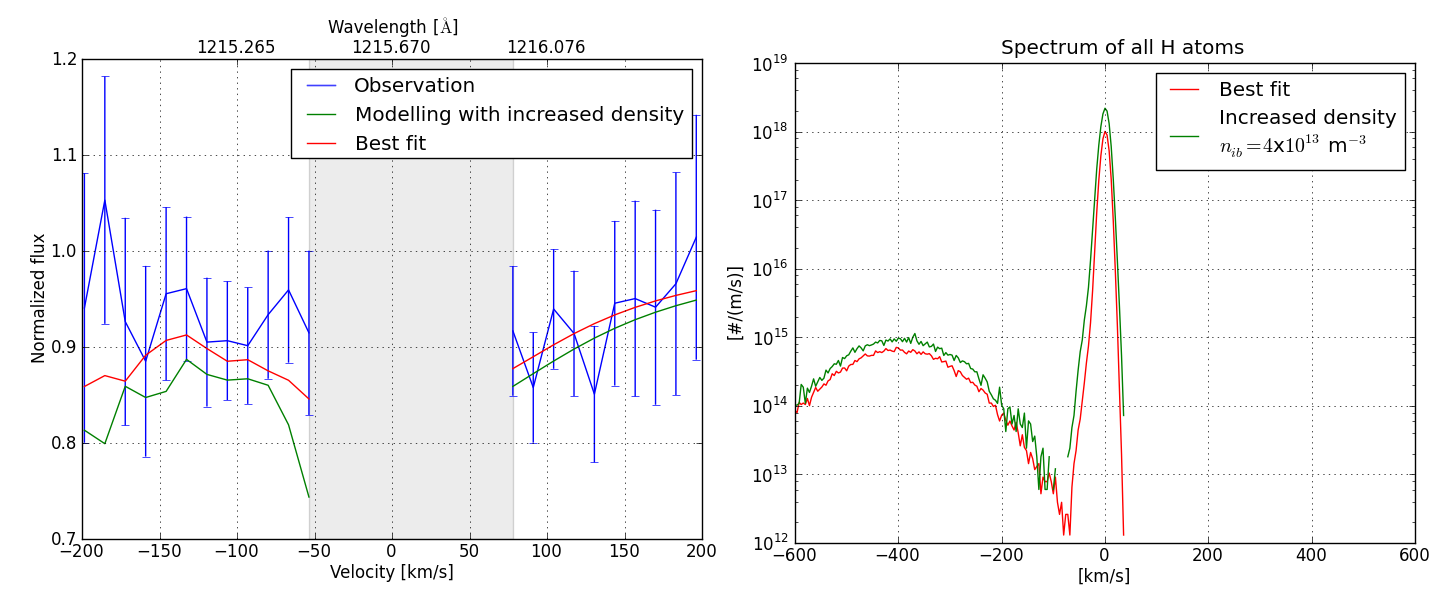}
  \caption{Left panel: the green line shows the best fit model, but with the inner boundary density increased by a factor of two from $2\times10^{13}$~m$^{-3}$ to $4\times10^{13}$~m$^{-3}$ ($\chi^2 \approx 0.181$). Right panel: the corresponding velocity spectra along the LOS.}
  \label{f_dens_Lya}
\end{figure}
%------------------------------------------------------------------

%------------------------------------------------------------------
\begin{figure}[t]
  \centering
  \includegraphics[width=1.0\textwidth]{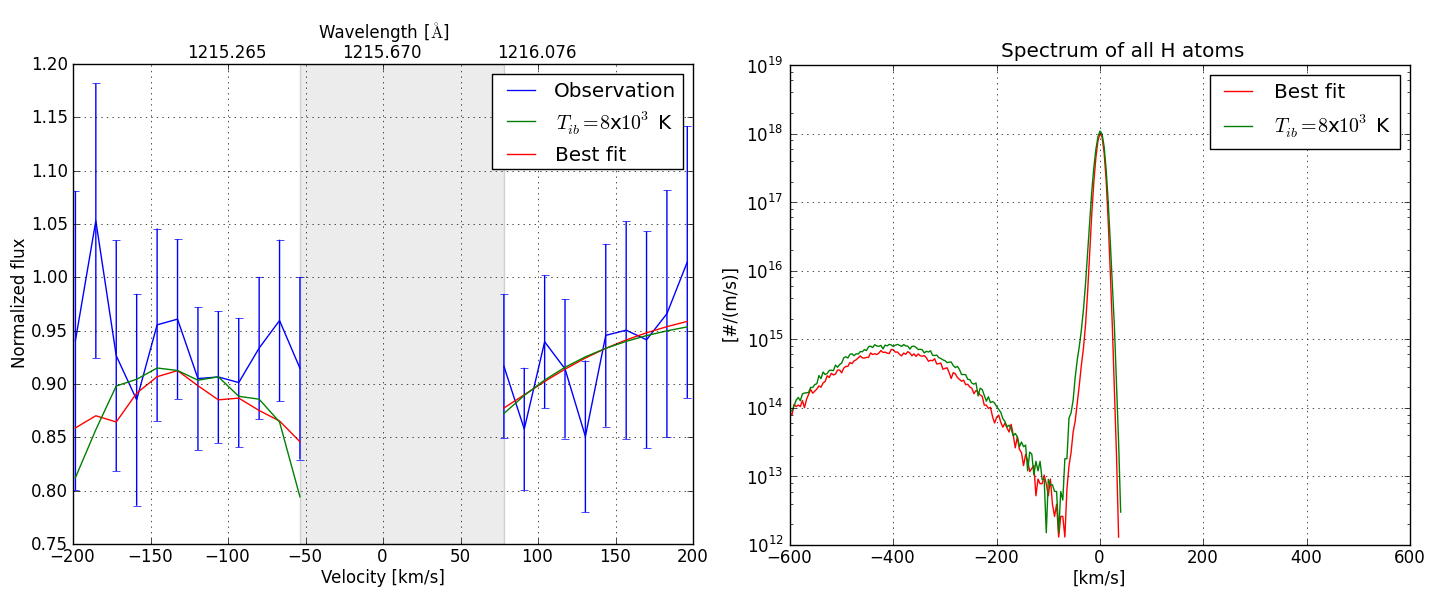}
  \caption{Left panel: the green line shows the best fit model, but with the inner boundary temperature of $8\times10^{3}$~K ($\chi^2 \approx 0.103$). Right panel: the corresponding velocity spectra along the LOS.}
  \label{f_temp_Lya}
\end{figure}
%------------------------------------------------------------------

%------------------------------------------------------------------
\begin{figure}[t]
  \centering
  \includegraphics[width=1.0\textwidth]{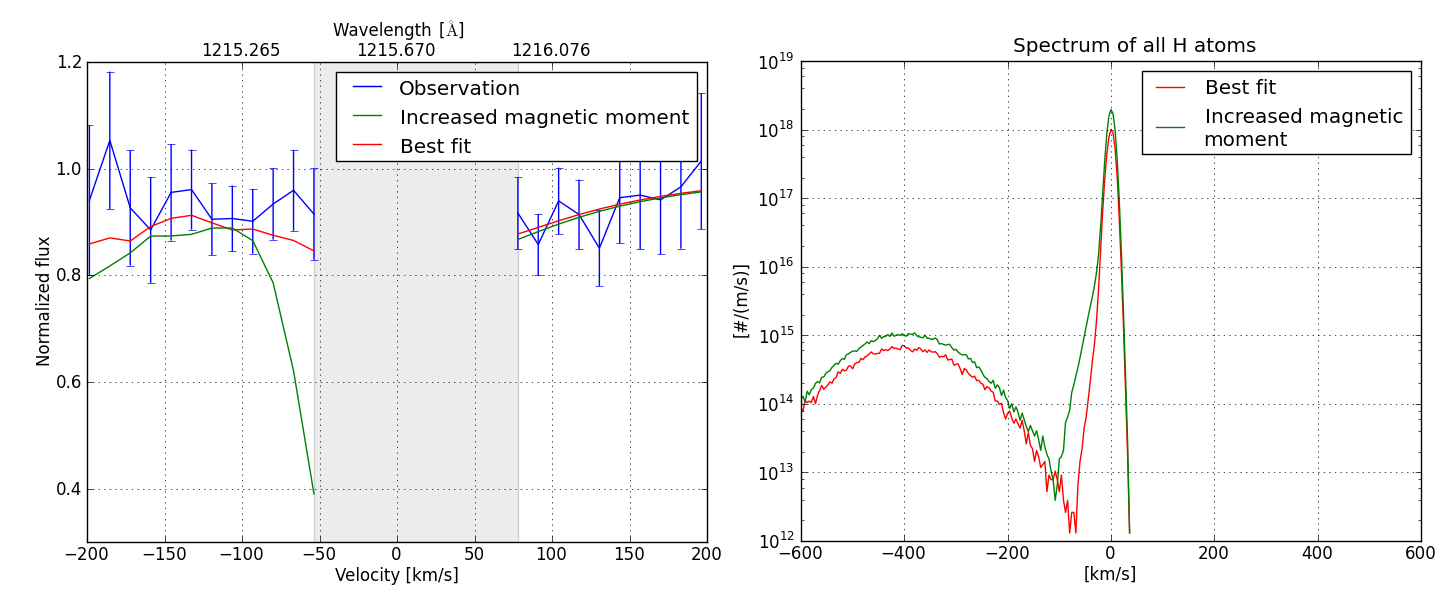}
  \caption{Left panel: the green line shows the best fit model, but with increased substellar distance $R_{\rm s} = 4$ and increased obstacle width $R_{\rm t} = 6$ ($\chi^2 \approx 0.558$). Right panel: the corresponding velocity spectra along the LOS.}
  \label{f_obst_Lya}
\end{figure}
%------------------------------------------------------------------

\begin{enumerate}
  \item \textbf{Broadening:} as discussed above, natural (and Doppler) broadening is the most important mechanism in the upper atmosphere of HD~209458b because of the low collision rate between neutral hydrogen atoms. This mechanism is the only one that can explain the presence of absorption also in the right wing of the Ly$\alpha$ line, which is in accordance with earlier work\cite{BJ10}. We can illustrate this by comparing in-transit spectra calculated with (Fig.~\ref{f_broad_Lya}, red line) and without (Fig.~\ref{f_broad_Lya}, green line) broadening. As one can see, without broadening, for the same velocity spectrum, the red wing of the modelled Ly$\alpha$ line shows practically no absorption, which contradicts the observation. Below we discuss separately density and temperature, which influence the absorption depth defined by broadening. As it was described above, we separately add the contribution from the lower atmosphere to the transmissivity.
  \item \textbf{Density:} density determines the influence of broadening (see Eqs.~\ref{e_T} and \ref{e_crossDelta}) and influences the depth of the absorption. A higher density also means presence of additional atoms undergoing acceleration by radiation pressure. Increasing the density leads to an increase of the optical depth, which in turn leads to a wider spectral line. Fig.~\ref{f_dens_Lya} illustrates the influence of density. As one can see, increasing the inner boundary density from $2 \times 10^{13}$~m$^{-3}$ by a factor of two to $4 \times 10^{13}$~m$^{-3}$ leads to significant overabsorption in the blue part of the modelled Ly$\alpha$ line. While the red part can be fitted inside the error bars using both densities, we conclude that a density of $2 \times 10^{13}$~m$^{-3}$ at $R_{\rm ib} = 2.8 R_{\rm pl}$ can better reproduce the observation in the blue wing. We conclude also that Ly$\alpha$ observations can be used to determine the density in exoplanetary upper atmospheres at the time of transit, since they can be fitted only in a narrow parameter range.
  \item \textbf{Temperature:} temperature influences the Doppler velocity spread, and, thus, broadening and changes the corresponding Ly$\alpha$ absorption in a similar way to density. Increasing the temperature leads to a corresponding increase in the absorption depth (Fig.~\ref{f_temp_Lya}). As illustrated in Fig.~\ref{f_temp_Lya}, the modelling is not as sensitive to the temperature change as it is to the density change. This leads us to the conclusion that the temperature at the time of observation was probably in the range of $6 \times 10^{3}~{\rm K}< T_{\rm ib} < 8 \times 10^3~{\rm K}$, which is in agreement with atmospheric models\cite{Koskinen13}. However, our modelling does not allow us to strictly constrain the upper atmosphere temperature.
  \item \textbf{Obstacle shape:} the shape of the magnetospheric obstacle influences the modelling results quite significantly. In the main part of the article we have stated that the modelling can be used to determine the magnetic properties of HD~209458b and to estimate its magnetic moment. Fig.~\ref{f_obst_Lya} illustrates the role played by the obstacle shape. As one can see, a bigger magnetosphere defined by a stronger magnetic field leads to a stronger magnetospheric protection of the atmosphere and, hence, to a corresponding increase of neutrals. The stronger magnetic field diminishes the production of ENAs and the electron impact ionization. Since these atoms undergo acceleration by the radiation pressure, an overabsorption in the blue wing of the Ly$\alpha$ arises (Fig.~\ref{f_obst_Lya}, green line) and the observation cannot be reproduced anymore. From the results of the modelling we conclude that HD~209458b has most likely a magnetic moment of about 10\% of $\mathcal{M}_{\rm Jup}$, or the magnetic moment of Jupiter, which is required to fit the observation.
  \item \textbf{Radiation pressure and self-shielding parameter:} it was shown\cite{BL13} that since the upper atmosphere of HD~209458b is optically thick to Ly$\alpha$, not all of the neutral hydrogen atoms in the vicinity of the planet can undergo acceleration by the radiation pressure, but only those located in the optically thin shell at the cloud's boundary. Our modelling is in agreement  with these results. For illustration we show the modelled hydrogen clouds (Fig.~\ref{f_shield}) and corresponding calculated Ly$\alpha$ in-transit and velocity spectra (Fig.~\ref{f_shield_Lya}). In the absence of self-shielding, the entire atmosphere above $R_{\rm ib}$ is accelerated, forming a very long cometary-like tail, which is ionized by the electrons of the stellar wind outside the magnetic obstacle. Inside the magnetosphere, the neutrals are ionized only by the stellar UV radiation, which is not strong enough to prevent overabsorption. The presence of such a large number of accelerated neutral hydrogen atoms leads to strong overabsorption and makes it complete impossible to reproduce the observation with any other parameters in a physically realistic range. 
  \item \textbf{Stellar wind density and velocity and the role of ENAs:} the influence of ENAs on the absorption is defined by several parameters. First, by the stellar wind velocity, which we assume to have a Maxwellian distribution to which the ENA distribution is proportional. This leads to the coincidence of the maxima in the velocity spectrum for these two distributions. Assuming a slow stellar wind shifts the ENAs population into the velocity domain of interest (restricted to $-200\le v_x \le200$~km/s, since the observation becomes too noisy beyond). Second, the stellar wind and exosphere density play a role defining the maximum possible number of ENAs that can form. A strong magnetic field provides an effective shielding of the upper atmosphere from the stellar wind electrons and prevents charge exchange with stellar wind protons. However, it increases also the number of neutrals accelerated by the radiation pressure and causes overabsorption (see discussion about magnetospheric properties above). Fig.~\ref{f_ENAs} illustrates the influence of the previously mentioned parameters on the modelled absorption spectrum. It shows the influence of decreased stellar wind velocity (green line), magnetic boundary (black line), and decreased stellar wind density and velocity (cyan line) on ENA production and the modelled spectrum. As one can see, for any of the considered parameter ranges, the observation cannot be reproduced if one assumes a slow stellar wind velocity. Fig.~\ref{f_cloud_ENAs} presents two corresponding hydrogen clouds around HD~209458b. The bigger tilt of the magnetic obstacle is caused by the smaller ratio of the stellar wind velocity to the orbital velocity of HD~209458b.
  \item \textbf{Numerical parameters, the role of the velocity grid and upsampling parameter:} as noted above, numerics plays a significant role and should be treated cautiously to avoid unphysical results. As illustration, we show in Fig.~\ref{f_num} three velocity and corresponding Ly$\alpha$ attenuation spectra calculated with best-fit physical parameters. The best fit was obtained with a fine velocity grid and high upsampling parameter (400). The green line in the left panel of Fig.~\ref{f_num} shows the calculation for a very low upsampling parameter (2). As one can see, the discontinuity of the cloud leads to a decrease of the absorption in the red part and increase in the blue part. The black line in Fig.~\ref{f_num} shows the Ly$\alpha$ transmissivity calculated for a coarse velocity grid (20 bins), whereas the best fit was calculated with binning to 301 bins. As one can see, a coarser grid leads to significant overabsorption and unphysical results. In our simulations, we have chosen the upsampling parameter of 400 as the one higher than the saturation value (i.e., the value above which the result stops being sensitive to changes in the upsampling parameter). The same is true for the number of velocity and spatial bins. However, the influence of the spatial grid is of less importance, and therefore a lower parameter can be chosen.
\end{enumerate}

%------------------------------------------------------------------
\begin{figure}[t]
  \centering
  \includegraphics[width=1.0\textwidth]{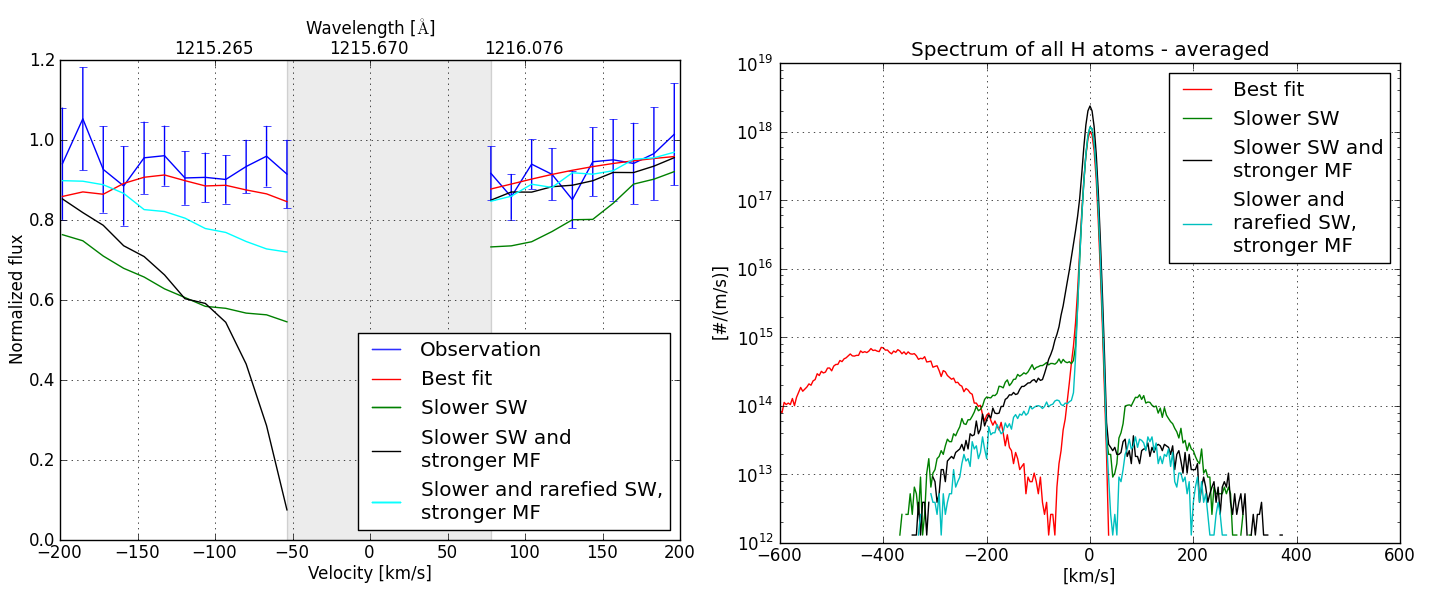}
  \caption{Illustration of the slower stellar wind with $v_{\rm sw} = 50$~km/s. Left panel: the green line shows the best fit model, but with $v_{\rm sw} = 50$~km/s ($\chi^2 \approx 1.383$). The magnetic obstacle is located at the same distance as in the best fit. The black line illustrates the model for $v_{\rm sw} = 50$~km/s and a big magnetosphere with $R_s=10 R_{\rm pl}$ and $R_t=15 R_{\rm pl}$ ($\chi^2 \approx 2.140$). The cyan line illustrates the influence of the the stellar wind density, which is decreased to $n_{\rm sw} = 10^9$~m$^{-3}$ ($\chi^2 \approx 0.267$). $v_{\rm sw} = 50$~km/s. Right panel: the corresponding velocity spectra along the LOS. The gap in the velocity spectrum with slow stellar wind near the main peak comes from the radiation pressure since the atoms in this velocity domain have the highest UV absorption rates (Fig.~\ref{f_rp}).}
  \label{f_ENAs}
\end{figure}
%------------------------------------------------------------------

%------------------------------------------------------------------
\begin{figure}[t]
  \centering
  \includegraphics[width=1.0\textwidth]{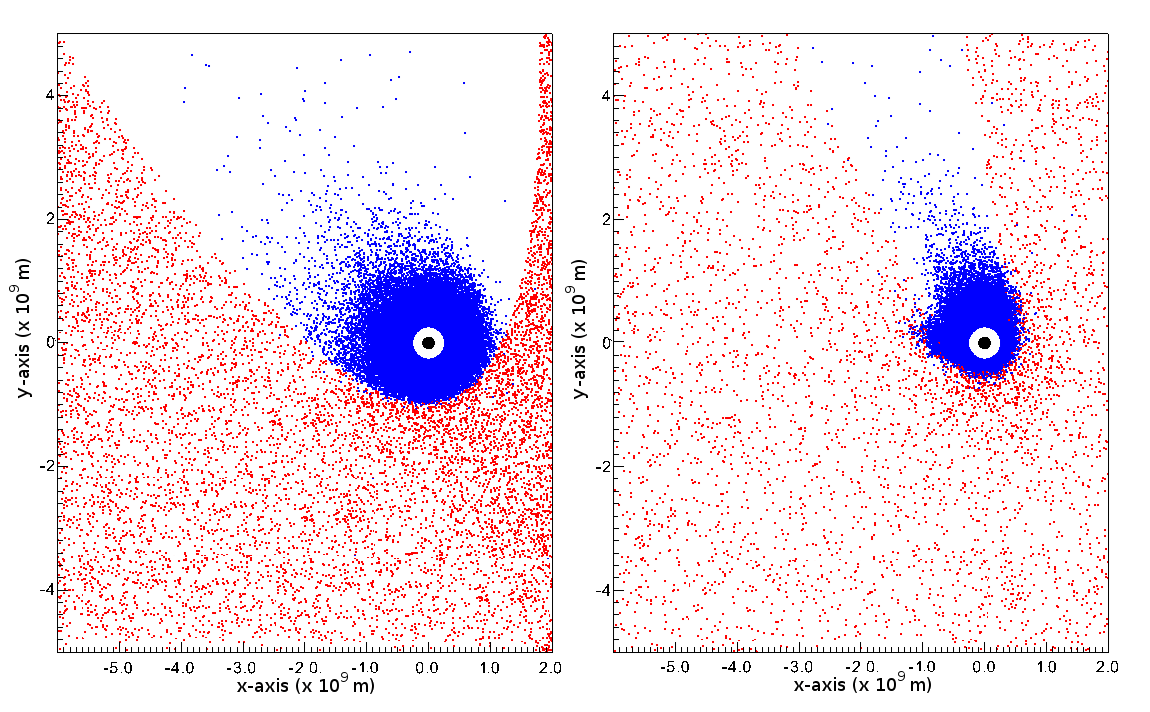}
  \caption{Comparison of modelled neutral hydrogen clouds around HD~209458b with slow stellar wind of 50 km/s. Presented are slices of modelled 3D atomic corona around the planet for $-10^7 \le z \le 10^7$~m. Blue and red dots correspond to neutral hydrogen atoms and hydrogen ions, which include stellar wind protons, respectively. The black dot in the center represents the planet. The white empty area around the planet corresponds to the XUV heated thermosphere up to the height $R_{\rm ib}$. Left panel: $R_{\rm s} = 10~R_{\rm pl}$, $R_{\rm t} = 15~R_{\rm pl}$ (for corresponding spectum see black line in Fig.~\ref{f_ENAs}). Right panel: additionally decreased stellar wind density of $n_{\rm sw} = 10^9$~m$^{-3}$ (see cyan line in Fig.~\ref{f_ENAs}). All other parameters are taken from the best fit.}
  \label{f_cloud_ENAs}
\end{figure}
%------------------------------------------------------------------

%------------------------------------------------------------------
\begin{figure}[t]
  \centering
  \includegraphics[width=1.0\textwidth]{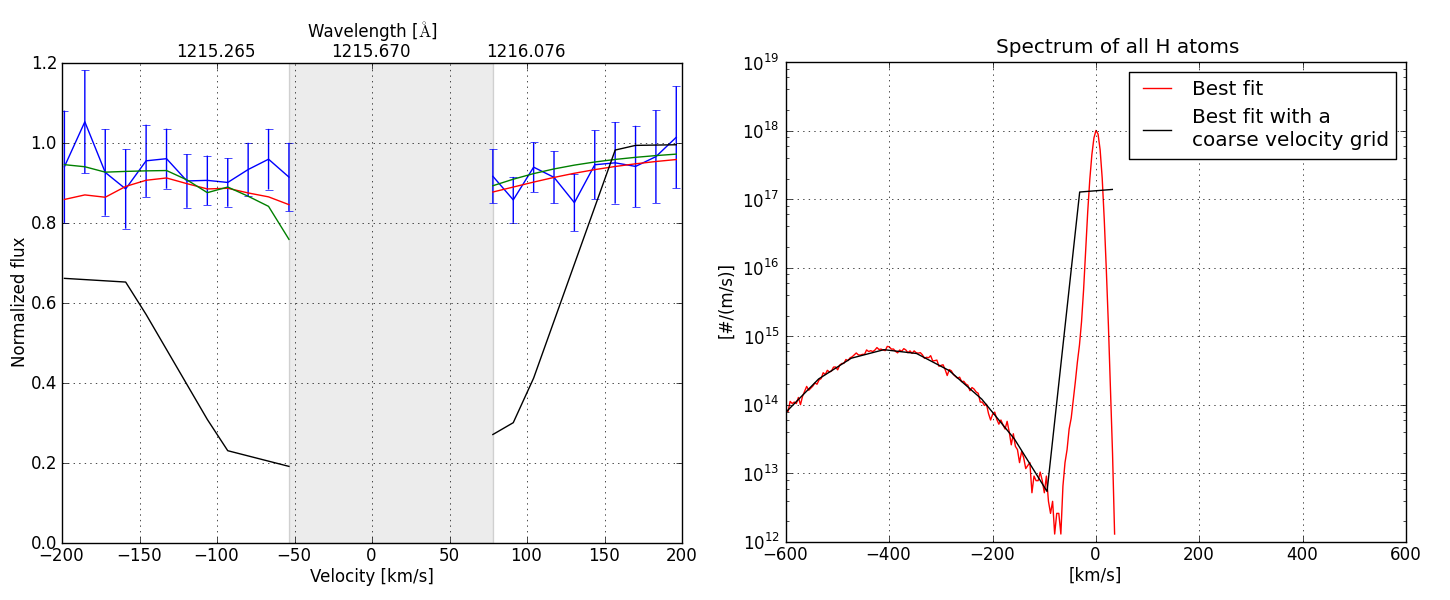}
  \caption{Left panel: the green line shows the best fit model, but calculated for a low upsampling parameter (2) and a fine velocity grid. The black line illustrates the result obtained assuming a low upsampling parameter (2) and a coarse velocity grid (20 bins instead of 301 used in the best fit), $\chi^2 \approx 4.954$. Right panel: the corresponding velocity spectra along the LOS for fine and coarse velocity grids.}
  \label{f_num}
\end{figure}
%------------------------------------------------------------------

% For your review copy (i.e., the file you initially send in for
% evaluation), you can use the {figure} environment and the
% \includegraphics command to stream your figures into the text, placing
% all figures at the end.  For the final, revised manuscript for
% acceptance and production, however, PostScript or other graphics
% should not be streamed into your compliled file.  Instead, set
% captions as simple paragraphs (with a \noindent tag), setting them
% off from the rest of the text with a \clearpage as shown  below, and
% submit figures as separate files according to the Art Department's
% instructions.

\end{document}